\documentclass[manuscript,screen]{acmart}

\AtBeginDocument{%
  }

\usepackage[colorinlistoftodos]{todonotes}
\usepackage{subcaption}
\usepackage{tablefootnote}
\usepackage{amsmath,amsfonts}
\usepackage{algorithmic}
\usepackage{algorithm}
\usepackage{array}
\usepackage{textcomp}
\usepackage{stfloats}
\usepackage{url}
\usepackage{verbatim}
\usepackage{graphicx}
\usepackage[shortlabels]{enumitem}

\usepackage{orcidlink}

\usepackage{multirow}
\usepackage{multicol}
\usepackage{ragged2e}
\usepackage{array}
\usepackage{makecell}
\usepackage{siunitx}
\usepackage[font=small]{caption}
\usepackage{xcolor}

\usepackage{adjustbox}
\usepackage{titlesec}

\usepackage{tikz}
\usetikzlibrary{shapes.geometric, arrows}
\tikzstyle{reference_rectangle} = [rectangle, align=center, draw=black, fill=white, minimum width=1cm, text width=2cm, font={\footnotesize}]
\tikzstyle{cat_rectangle} = [draw, rectangle, rounded corners, align=center, draw=black, fill=white, minimum width=2cm, text width=2cm, font={\footnotesize}]
\tikzstyle{counter_rectangle} = [draw, rectangle, rounded corners, align=center, draw=black, fill=black!10, minimum width=2cm, text width=2cm, font={\footnotesize}]
\tikzstyle{label} = [draw, rectangle, align=center, draw=black, fill=white, minimum width=0.2cm, text width=0.7cm, font={\footnotesize}]
\tikzstyle{point} = [font=\bfseries]
\tikzstyle{attack_components} = [draw, rectangle, rounded corners, align=center, draw=black, fill=white, minimum width=2cm, minimum height=1cm, text width=2cm, font={\footnotesize}]
\tikzstyle{dff} = [draw, rectangle, align=center, draw=black, fill=white]

\definecolor{redish}{HTML}{F8BABA}
\definecolor{blueish}{HTML}{D9E6FC}
\definecolor{greenish}{HTML}{C8E8C5}

\tikzstyle{arrow} = [thick,->,>=stealth]
\usepackage[nolist]{acronym}

\begin{acronym}
    \acro{EIST}{Enhanced Intel Speedstep Technology}
    \acro{OPP}{Operating Point}
    \acro{CPU}{Central Processing Unit}
    \acro{MSR}{Model-Specific Register}
    \acro{DFA}{Differential Fault Analysis}
    \acro{DVFS}{Dynamic Voltage and Frequency Scaling}
    \acro{TEE}{Trusted Execution Environment}
    \acro{REE}{Rich Execution Environment}
    \acro{ISA}{Instruction Set Architecture}
    \acro{TPM}{Trusted Platform Module}
    \acro{OS}{Operating System}
    \acro{TA}{Trusted Applications}
    \acro{SE}{Secure Element}
    \acro{SGX}{Software Guard Extension}
    \acro{SEV}{Secure Encrypted Virtualization}
    \acro{MMU}{Memory Management Unit}
    \acro{SE}{Secure Element}
    \acro{QSEE}{Qualcomm Secure Execution Environment}
    \acro{TCB}{Trusted Computing Base}
    \acro{RoT}{Root of Trust}
    \acro{NS}{Non-Secure}
    \acro{SCR}{Secure Configuration Register}
    \acro{SM}{Secure Monitor}
    \acro{EPC}{Enclave Page Cache}
    \acro{EPCM}{Enclave Page Cache Map}
    \acro{PRM}{Processor Reserved Memory}
    \acro{AEX}{Asynchronous Enclave Exit}
    \acro{LLC}{Last-Level Cache}
    \acro{TLB}{Translation Lookaside Buffer}
    \acro{PTW}{Page Table Walker}
    \acro{PMP}{Physical Memory Protection}
    \acro{MPU}{Memory Protection Unit}
    \acro{SCA}{Side-Channel Attack}
    \acro{DFA}{Differential Fault Analysis}
    \acro{DoS}{Denial-of-Service}
    \acro{FIA}{Fault Injection Attack}
    \acro{PMIC}{Power Management Integrated Circuit}
    \acro{RAPL}{Running Average Power Limit} 
    \acro{KASLR}{Kernel Address Space Layout Randomization}
    \acro{UI}{User Interface}
    \acro{SoC}{System on Chip}
    \acro{OS}{Operating System}
    \acro{PoC}{Proof-of-Concept}
    \acro{PLL}{Phase-Locked Loop}
    \acro{IVR}{Integrated Voltage Regulator}
    \acro{FPGA}{Field Programmable Gate Array}
    \acro{PP}{Protection Profile}
    \acro{ICMP}{Internet Control Message Protocol}
    \acro{AVS}{Adaptive Voltage Scaling}
    \acro{PDN}{Power Distribution Network}
    \acro{DFF}{D-Flip-Flop}
    \acro{AES}{Advanced Encryption Standard}
\end{acronym}
\newcommand{\countermeasureSubsubsection}[1]{
  \titleformat{\subsubsection}
    {\normalfont\normalsize\bfseries}
    {}
    {1em}
    {}
  \subsubsection{#1}
  \titleformat{\subsubsection}
    {\normalfont\normalsize\itshape}
    {\arabic{subsubsection})}
    {1em}
    {}
}

\newcommand{\countermeasureParagraph}[1]{
  \titleformat{\paragraph}
    {\normalfont\normalsize\itshape}
    {}
    {1em}
    {}
  \paragraph{#1}
  \titleformat{\paragraph}
    {\normalfont\normalsize\itshape}
    {\alph{paragraph})}
    {1em}
    {}
}
\begin{document}

\title{Do Not Trust Power Management: A Survey on Internal Energy-based Attacks Circumventing Trusted Execution Environments Security Properties}

\author{Gwenn Le Gonidec}
\email{owen.legonidec@univ-nantes.fr}
\orcid{0009-0004-4774-6886}
\affiliation{
    \institution{UnivRennes, INSA Rennes, CNRS, IETR-UMR 6164}
    \city{Rennes}
    \country{France}
}

\author{Maria M\'endez Real}
\email{maria.mendez-real@univ-ubs.fr}
\orcid{0000-0002-9336-9936}
\affiliation{
    \institution{UMR 6285, Lab-STICC, Univ. Bretagne-Sud}
    \city{Lorient}
    \country{France}
}

\author{Guillaume Bouffard}
\email{guillaume.bouffard@ssi.gouv.fr}
\orcid{0000-0002-2046-369X}
\affiliation{
    \institution{National Cybersecurity Agency of France (ANSSI)}
    \city{Paris}
    \country{France}
}

\author{Jean-Christophe Pr\'evotet}
\email{jean-christophe.prevotet@insa-rennes.fr}
\orcid{0000-0001-6951-4702}
\affiliation{
    \institution{UnivRennes, INSA Rennes, CNRS, IETR-UMR 6164}
    \city{Rennes}
    \country{France}
}

% --------------------------------------------------------------------

\begin{abstract}

Over the past few years, several research groups have introduced innovative hardware designs for Trusted Execution Environments (TEEs), aiming to secure applications against potentially compromised privileged software, including the kernel~\cite{costanSanctumMinimalHardware2016a,pintoDemystifyingArmTrustZone2019}.
Since 2015~\cite{yanStudyPowerSide2015}, a new class of software-enabled hardware attacks leveraging energy management mechanisms has emerged. These internal energy-based attacks comprise fault~\cite{tangCLKSCREWExposingPerils2017}, side-channel~\cite{lippPLATYPUSSoftwarebasedPower2021} and covert channel attacks~\cite{haj-yahyaIChannelsExploitingCurrent2021}. Their aim is to bypass TEE security guarantees and expose sensitive information such as cryptographic keys. They have increased in prevalence in the past few years~\cite{koglerCollidePowerLeaking2023,chowdhuryyPowSpectrePoweringSpeculation2024,grossFPGANeedlePreciseRemote2023}.
Popular TEE implementations, such as ARM TrustZone and Intel SGX, incorporate countermeasures against these attacks. However, these countermeasures either hinder the capabilities of the power management mechanisms or have been shown to provide insufficient system protection~\cite{mishraTooHotHandle2024,chowdhuryyPowSpectrePoweringSpeculation2024}.
This article presents the first comprehensive knowledge survey of these attacks, along with an evaluation of literature countermeasures. We believe that this study will spur further community efforts towards this increasingly important type of attacks.

\end{abstract}

\maketitle

\acresetall{}

% --------------------------------------------------------------------

\section{Introduction}
To protect sensitive assets, modern mobile computing systems rely on \acp{TEE}. \acp{TEE} offer a secure runtime execution environment where sensitive applications are executed, co-located with a \ac{REE}, which may include modern and complex \acp{OS} such as Android or iOS. \acp{TEE} isolate the execution of sensitive applications from the rest of the system, \emph{i.e.}, from the untrusted \ac{REE}, relying on hardware mechanisms. \acp{TEE} are widely used to secure critical applications in \acp{SoC} embedded in powerful and complex devices, from smartphones to high-end servers. These \acp{SoC} are powered by complex \acp{CPU} with advanced micro-architectures, including memory virtualization, multiple cache levels, and speculative or out-of-order execution. \ac{TEE} and \ac{REE} are executed in the one \ac{CPU} and may share the same core. 
Designing innovative hardware-based protection mechanisms for next-generation \acp{TEE} is an active research field. Over the past few years, several research groups have leveraged the capabilities of the RISC-V open \ac{ISA} for this purpose~\cite{costanSanctumMinimalHardware2016a,leeKeystoneOpenFramework2020,schrammelSPEARVSecurePractical2023}.

\acp{TEE} aim to protect the system against attacks where the attacker may have full control over the \ac{REE}, including the \ac{OS}. %Attacks requiring physical access to the system are not part of the \ac{TEE} threat model
They typically do not provide protection against an attacker who has physical access to the device~\cite{pintoDemystifyingArmTrustZone2019,feiSecurityVulnerabilitiesSGX2022,dessoukyEnclaveComputingRISCV2020}. However, since the mid-2010s, numerous studies have demonstrated various methods by which a remote attacker can compromise the main security properties of a \ac{TEE}, utilizing hardware attacks initiated from software without needing physical access to the target~\cite{grussRowhammerJsRemote2016}. They rely on the manipulation of software-accessible interfaces to affect hardware components. This emerging class of attacks stems from the complexity of the systems on which TEEs are implemented, which attempt to balance performance, power constraints and security. At the intersection of software and hardware attacks, they benefit from the best of both approaches. They take advantage of hardware attack techniques developed over the last two decades and, as software attacks, they can be executed remotely and on a massive scale. This makes them a realistic and significant threat.

Of particular concern are attacks that maliciously exploit embedded power management mechanisms, referred to as \emph{internal energy-based attacks} in this survey. These attacks can be employed to extract sensitive assets such as cryptographic keys or to bypass authentication procedures. They enable the injection of timing faults in a \ac{FIA}~\cite{tangCLKSCREWExposingPerils2017, murdockPlundervoltSoftwarebasedFault2020}, extraction of secret information through a \ac{SCA}~\cite{diptaDFSCADynamicFrequency2022, wangHertzbleedTurningPower2023, lippPLATYPUSSoftwarebasedPower2021}, and the creation of a covert communication channel utilized by a Trojan and a spy~\cite{benhaniDVFSSecurityFailure2018, haj-yahyaIChannelsExploitingCurrent2021}. Such attacks have been conducted against the most widely deployed \ac{TEE} designs, namely Arm TrustZone~\cite{benhaniDVFSSecurityFailure2018,wangHertzbleedTurningPower2023,tangCLKSCREWExposingPerils2017} and Intel \ac{SGX}~\cite{murdockPlundervoltSoftwarebasedFault2020,mantelHowSecureGreen2018,lippPLATYPUSSoftwarebasedPower2021}.

Among commercial \ac{TEE} implementations, Intel \ac{SGX} has introduced a mitigation for energy-based \acp{FIA}. However, this approach involves restricting power management mechanisms, significantly limiting the primary objective of fine-grained energy optimization~\cite{murdockPlundervoltSoftwarebasedFault2020}. Arm~\cite{armPowerPerformanceManagement2019} recommends that vendors of TrustZone-based \acp{TEE} (e.g., Qualcomm or Samsung) implement similar mitigation; however, to our knowledge, no official vendor documentation confirms an actual implementation. This mitigation is ultimately unsustainable, as it leads to power waste and fails to address the root cause of energy-based attacks. For energy-based \acp{SCA}~\cite{wangHertzbleedTurningPower2023} and covert attacks, Intel has implemented a noise-based countermeasure~\cite{intelRunningAveragePower2022}. However, recent studies have demonstrated that it does not fully prevent such attacks~\cite{chowdhuryyPowSpectrePoweringSpeculation2024}. 
RISC-V-based \ac{TEE} designs do not embed any protection against internal energy-based attacks \cite{weiserTIMBERVTagIsolatedMemory2019,leeKeystoneOpenFramework2020,bahmaniCURESecurityArchitecture2021,schrammelSPEARVSecurePractical2023}.

Consequently, it is argued that existing \acp{TEE} are vulnerable to this significant threat. A promising area of research is the development of new countermeasures that can effectively counter internal energy-based attacks while preserving the full capabilities of power management mechanisms. 

This article provides a comprehensive overview of this new category of hardware energy-based attacks, analysing their capabilities, limitations, and evolution over the past few years. To the best of our knowledge, this is the first work to review this emerging category of attacks. Additionally, the initial attempts at countermeasures are reviewed. 

The remainder of this article is organized as follows: Section~\ref{sec:background} provides background information on physical attacks, power management mechanisms, and \acp{TEE}. Subsequently, Section~\ref{sec:attacks} delves into internal energy-based attacks, detailing their methods, results, and limitations. In Section~\ref{sec:discussion}, the first published and implemented countermeasures against energy-based attacks are scrutinized, highlighting their shortcomings and providing an overview of the challenges involved in their practical implementation. Finally, Section~\ref{sec:conclu} concludes this article and offers insights into future research directions.

% --------------------------------------------------------------------

\section{Background}
\label{sec:background}

\subsection{\aclp{TEE}} \label{sec:tees}
Some privileged components, such as \ac{REE}, have extensive attack surfaces that make them challenging to secure. For example, a rich \ac{OS} as Linux kernel, comprised over 27 million source lines of code in 2020, with nearly two thousand vulnerabilities disclosed~\cite{shameli-sendiUnderstandingLinuxKernel2021}. Therefore, these privileged components cannot be relied upon to ensure the security of critical programs. This realization gave rise to the concept of relying on a minimal set of security-oriented components to form a \ac{TCB}. The aim is to keep the \ac{TCB} as small as possible, providing hardware-assisted mechanisms to isolate critical programs from the main computing environment. Since the late 1990s, separate co-processors known as \acp{TPM} have been utilized for this purpose~\cite{osbornTrustedPlatformModule2013}.  However, programs running on a \ac{TPM} cannot benefit from the full power of the SoC's components. The necessity to extend this protection to third-party programs on rich, performance-oriented processors led to the development and standardization of \acp{TEE} in the early 2010s~\cite{sabtTrustedExecutionEnvironment2015}.

A \ac{TEE} is a secure and performance-oriented environment comprising memory, storage, and processing capabilities, isolated from the rest of the system, often referred to as the \ac{REE}~\cite{ekbergTrustedExecutionEnvironments2013}. While a \ac{TPM} relies on executing security-critical programs on an external, isolated component, in a \ac{TEE}, both trusted and untrusted programs run on the same \ac{CPU} and share the same hardware resources. These resources' access is typically mediated by a dedicated component known as the \textit{Secure Monitor}. The \ac{TEE} utilizes on-chip hardware mechanisms to isolate and provide integrity and confidentiality to security-critical programs in systems where privileged software, such as kernels and hypervisors in the \ac{REE}, is untrusted. This includes servers used by stakeholders for data storage and computation outsourcing (cloud computing), where trust in the cloud provider may be lacking. This scenario is a common use case for Intel \ac{SGX} and AMD SEV, which are proprietary \ac{TEE} designs. In the embedded market, Arm TrustZone provides hardware support for \acp{TEE} to \ac{SoC} designers. In the embedded world, \acp{TEE} can be used for various applications as well, such as secure telemetry, biometry and digital rights management. More recently, numerous research papers have leveraged the open RISC-V \ac{ISA} to propose new hardware designs for \ac{TEE} support, such as Sanctum~\cite{costanSanctumMinimalHardware2016a} and Keystone~\cite{leeKeystoneOpenFramework2020}, among others.

\begin{figure}
    \centering
    \includegraphics[width=0.4\linewidth]{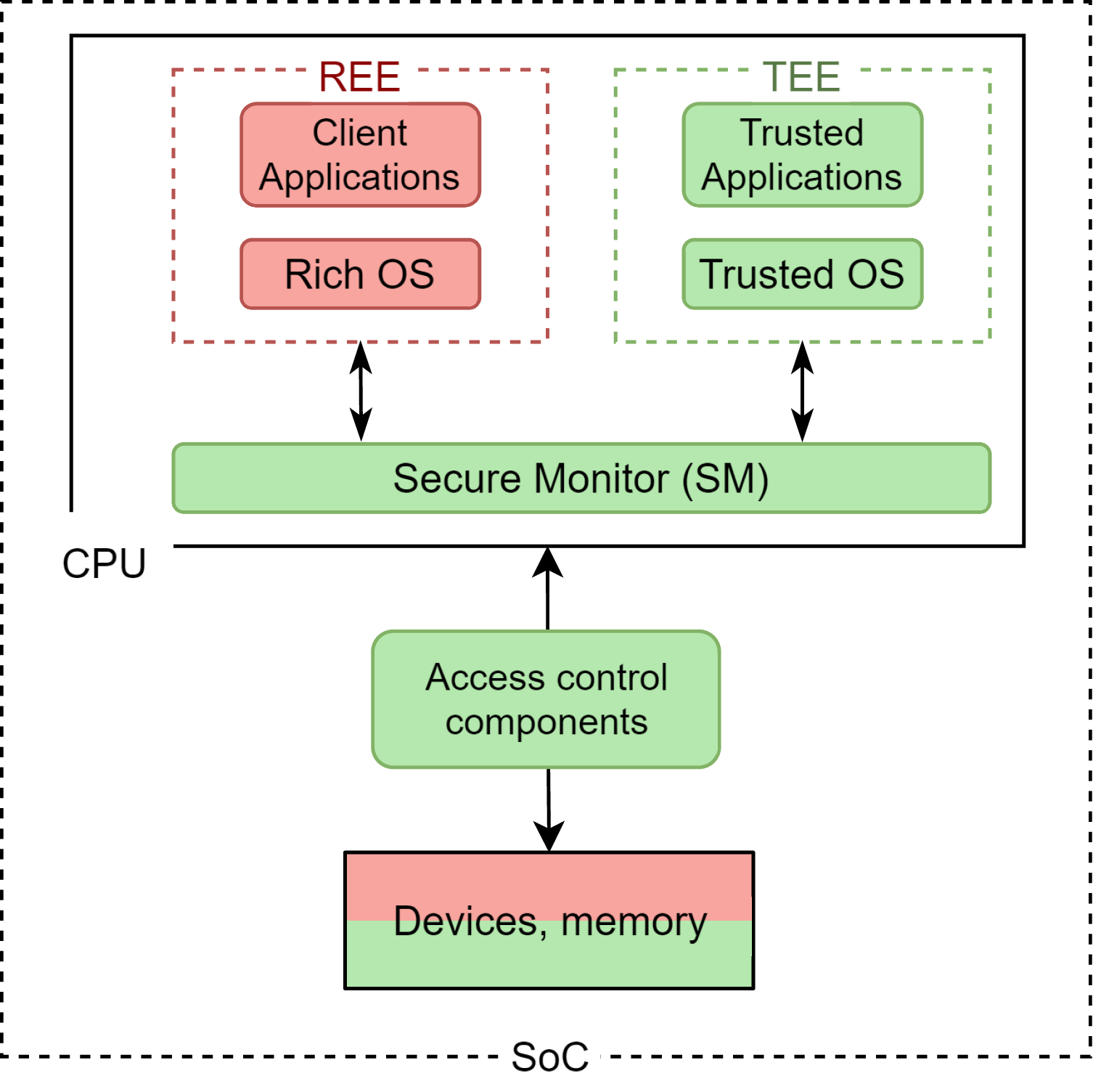}
    \caption{Example implementation of the main components in a dual-world-based TEE}
    \label{fig:tee}
\end{figure}

\ac{TEE} implementations vary based on the hardware and software mechanisms they employ to secure trusted applications. A notable distinction between Arm TrustZone and Intel \ac{SGX}-type \acp{TEE}, which encompass most academic proposals~\cite{costanSanctumMinimalHardware2016a,bahmaniCURESecurityArchitecture2021,schrammelSPEARVSecurePractical2023}, is that in the former, the entire system is divided into two \emph{worlds}; while the latter aims to secure individual applications known as~\emph{enclaves}. Figure~\ref{fig:tee} shows a simple implementation of such a dual-world based TEE, similar to TrustZone~\cite{pintoDemystifyingArmTrustZone2019}. Secure \acp{OS} and potentially hypervisors operate in the trusted world. In addition, the Secure Monitor may be able to reserve certain peripheral devices and areas of memory for the exclusive use of the trusted world. This approach offers significant flexibility: \ac{TEE} implementations can prioritize providing extensive functionalities to trusted applications or aim for the smallest possible \ac{TCB}, as demonstrated in studies~\cite{pintoDemystifyingArmTrustZone2019}. However, in the latter approach, untrusted user-level applications, referred to as enclaves, can be individually isolated without needing to trust the kernel and other privileged software. In this protection model, there is a risk of malicious enclaves being spawned.

Another essential feature of \ac{TEE} is attestation, which ensures that it operates on a physically certified device. This serves various purposes, such as allowing a trusted application to verify that the host system is running the latest version. Additionally, \acp{TEE} rely on security primitives like Secure Boot, typically enforced by firmware, which forms part of the \ac{TCB} of the system. 

Recently, an important area of research has emerged in the design of innovative hardware mechanisms for next-generation TEE. They leverage RISC-V's capabilities to create new TEE designs with various objectives, such as adaptation to real-time constraint~\cite{weiserTIMBERVTagIsolatedMemory2019}, flexibility~\cite{bahmaniCURESecurityArchitecture2021} and robustness against advanced attack schemes, e.g. controlled-channel attacks relying on page faults~\cite{schrammelSPEARVSecurePractical2023}.

\subsection{Power Management Mechanisms} \label{sec:dvfs}

Optimal power management is crucial in digital systems design, particularly for complex \acp{CPU}, enabling energy savings and improved thermal management. Among power management technologies, fine-grained control mechanisms such as \ac{DVFS} have been employed for decades. \ac{DVFS} involves adjusting the supply voltage and operating frequency, which together have a cubic impact on power consumption based on system load to meet performance targets. In this article, a combination of operating frequency and supply voltage is referred to as an \ac{OPP} (known as \emph{P-States} in Intel architecture terminology). In a typical \ac{DVFS} implementation, either the \ac{OS} or the hardware \ac{CPU} requests a frequency adjustment based on workload estimation. The operating voltage is subsequently adjusted based on a set of vendor-defined \acp{OPP}. These sets are defined by manufacturers to ensure device safety and can extend beyond the actual operating limits of the device~\cite{zhangBlacklistCoreMachineLearning2018}. These limits vary due to factors such as ageing, temperature, and manufacturing uncertainties~\cite{tangCLKSCREWExposingPerils2017}, making it impractical for operators to measure the limits of each device. In addition to \acp{OPP} scaling mechanisms, the ability to query power consumption, temperature, and operating frequency is crucial for power management. Energy-aware programs benefit from software interfaces that allow them to monitor these metrics, and it is also directly used by hardware components to implement turbo frequencies and power capping~\cite{schoneEnergyEfficiencyAspects2021}.

\begin{figure}[!hbt]
    \centering
    \includegraphics[width=0.5\linewidth]{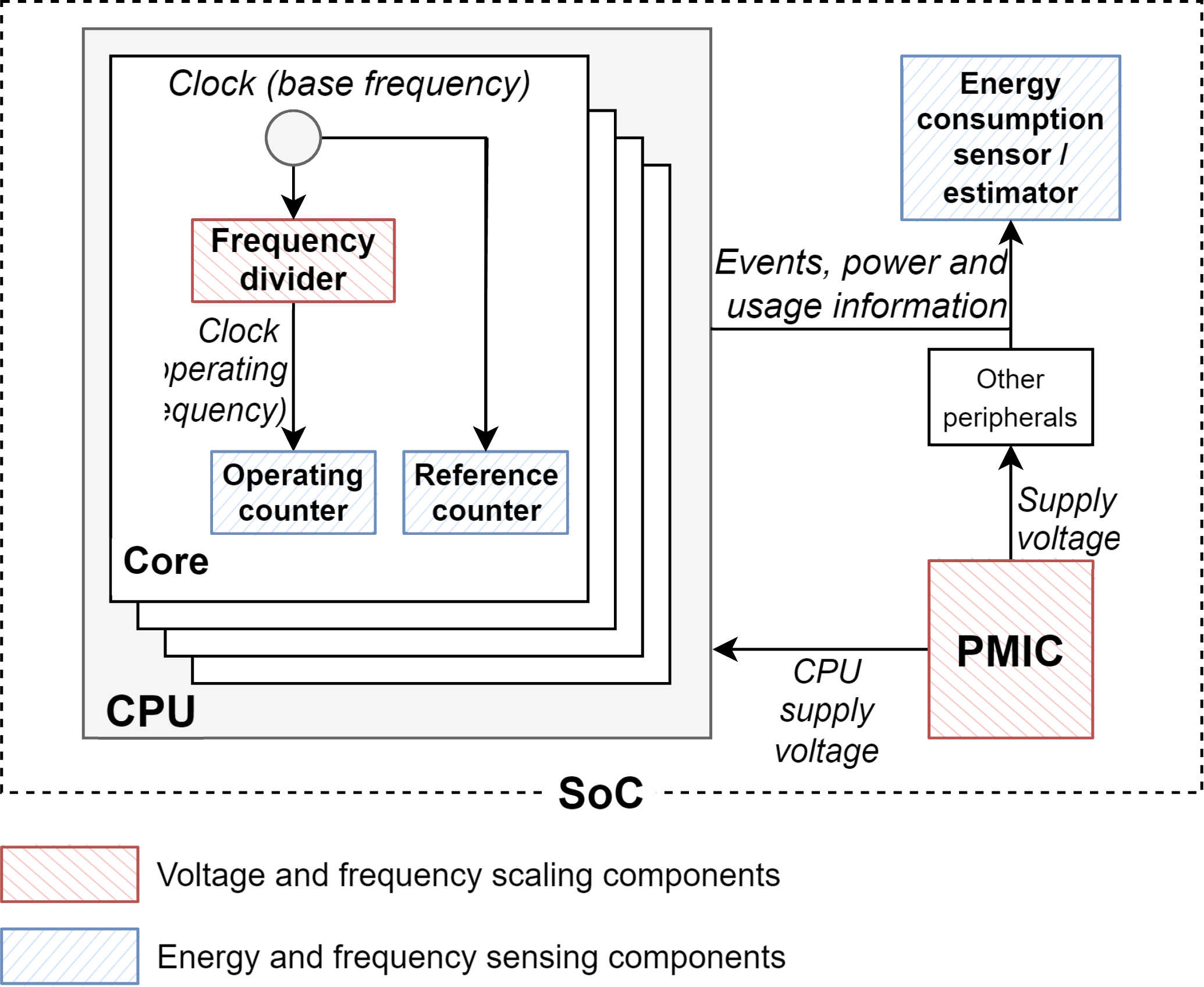}
    \caption{\small A \ac{DVFS} implementation in a SoC, illustrating typical components used for sensing operating frequency and supply voltage.}
    \label{fig:power-components}
\end{figure}

Figure~\ref{fig:power-components} provides an overview of the main components utilized in a typical \ac{DVFS} implementation within a \ac{SoC} featuring an application multi-core \ac{CPU}, both for \ac{OPP} scaling and sensing. It represents one implementation among many possibilities. Indeed, hardware implementations of such power management mechanisms can vary significantly from vendor to vendor and are often undocumented.

Energy consumption metrics can be retrieved by various means, depending on the platform considered. For instance, in Android-based devices, the power supply module communicates battery status to privileged software. On x86 processors, energy consumption can be measured from software using the \ac{RAPL} interface. Depending on the context, this interface does not always rely on physical power sensors; it may instead utilize a model based on architectural events communicated by peripherals~\cite{bosePowerManagementMulticore2012}. 

The operating frequency is correlated to power consumption, as both are tied by \ac{DVFS} mechanisms. It is typically measured by differentiating the values of incrementing counters; one incrementing with a reference frequency and the other incrementing in proportion to the actual performance\footnote{See: \texttt{aperfmperf.c} in the Linux kernel source code.}~\cite{intelIntel64IA322021}.

Frequency regulation in a \ac{CPU} is handled by frequency dividers using \acp{PLL}. Historically, voltage regulation has been predominantly managed by an off-chip~\ac{PMIC}, as represented in Figure~\ref{fig:power-components}. However, on-chip \ac{IVR} are becoming increasingly common~\cite{bairamkulovPlacementOnchipDistributed2023}, allowing for fast voltage transitions. Power is brought to the SoC components by a shared \ac{PDN}. At the software level, interaction with the frequency and voltage regulators is facilitated through dedicated OS modules. For instance, in Linux-based systems, the \ac{OS} kernel requests frequency changes through modules such as \texttt{cpufreq}, either using an automatic governor or manually setting an operating frequency through \acp{MSR}. Voltage and current regulators can be controlled via a dedicated framework~\cite{linuxkerneldocumentationRegulatorConsumerDriver2008}.

High voltage is required to achieve high operating frequencies. If the voltage is insufficient, timing conditions at the transistor level may not be met, resulting in device instability and clock glitches~\cite{gominaPowerSupplyGlitch2014}. Although supply voltage is typically adjusted based on operating frequency, it is possible to independently adjust them by directly writing to dedicated \acp{MSR}. This presents an entry point for internal energy-based \ac{FIA}, which intentionally triggers clock glitches by configuring a high clock frequency and a low supply voltage.

Finally, energy management mechanisms are evolving, becoming more aggressive and sophisticated. While a core cluster used to be typically supplied by a single power rail, per-core supply voltage domains are becoming more common, as implemented in some recent Intel architecture-based \ac{CPU}~\cite{schweikhardtDFSMixedCriticality2022}. Additionally, the time required to change supply voltage is decreasing. The first energy-based \ac{FIA} were conducted on systems where a change in voltage took about a millisecond to complete~\cite{murdockPlundervoltSoftwarebasedFault2020}. Nowadays, \acp{IVR} with fast voltage transitions are increasingly common. Intel's hardware-managed performance states significantly reduce the latency of both frequency and voltage changes~\cite{schweikhardtDFSMixedCriticality2022,deveyPowerManagementTechnology2021}. The literature demonstrates \acp{IVR} with sub-microsecond voltage transition delays~\cite{kellerRISCVProcessorSoC2017}. Therefore, one can expect energy management mechanisms to continue evolving in this direction with faster, finer, more sophisticated energy management strategies.

\subsection{Hardware Attacks} \label{sec:physical-attacks}
Hardware attacks exploit the electronic behaviour of components executing sensitive or critical applications. In the case of \ac{SCA}, rather than breaking complex mathematical algorithms such as cryptographic schemes protecting secret data, an attacker analyses side-channel signals generated during the execution of the victim application. Examples include the target device's electromagnetic emanation~\cite{camuratiScreamingChannelsWhen2018}, power consumption~\cite{mangardPowerAnalysisAttacks2008}, or computation time~\cite{geSurveyMicroarchitecturalTiming2018}. These signals reveal the target architecture and microarchitecture state, allowing an attacker to deduce the instructions and/or data computed, thereby compromising target confidentiality. 
Similarly, in \acp{FIA}, an attacker can introduce faults into computation or memory by inducing glitches within the electronic device through external sources, potentially resulting in controlled instruction skipping, such as during an authentication process.
In cryptographic application, an attacker may fault the result of specific instructions to infer cipher keys using dedicated methods such as \ac{DFA}.
In both cases, \ac{SCA} or \ac{FIA}, the objective is typically to retrieve secret information.

Traditionally, hardware attacks have required physical access to the hardware device, such as through external sensors or sources of glitches. Consequently, hardware attacks were not always considered in most device threat models. However, a new type of software-induced hardware attack has emerged, allowing an attacker application to perform hardware attacks from within the device. By exploiting microarchitectural components shared by the victim and attacker applications, such as memory caches~\cite{saxenaCacheBasedSide2018}, communication interconnections~\cite{sepulvedaCacheAttacksExploiting2021}, or more recently, power management components, this powerful new type of attack greatly extends the threat surface.

\subsection{Timing constraint in digital systems} \label{sec:timing-constraints}

\begin{figure}
    \centering
    \includegraphics[width=0.65\linewidth]{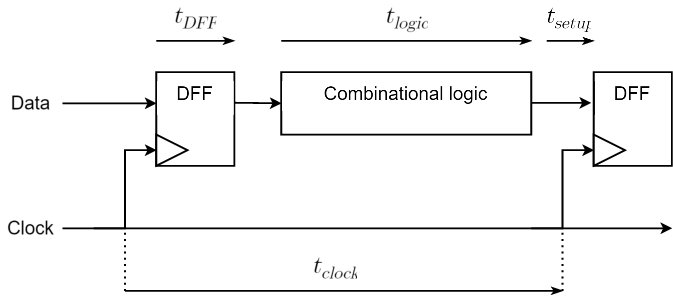}
    \caption{Timing constraints in a digital system}
    \label{fig:timing-constaints}
\end{figure}

Digital systems are ruled by timing constraints. As shown in Figure~\ref{fig:timing-constaints}, a synchronous circuit (such as an instruction) typically consists of two \acp{DFF}, which value changes upon receiving the synchronizing clock signal every $t_{clock}$. In-between those two is the combinational logic of the instruction. The output logic signal needs to be constant for $t_{setup}$ during the clock signal's receiving in order for the output DFF to correctly register it. The input signal also needs to be constant for $T_{DFF}$ after receiving the clock signal. Therefore, the combinational logic has to generate the output signal within $t_{logic}$, this time being ruled by the Equation~\ref{eq:timing-constaints}:

\begin{equation}
    t_{logic} \leq t_{clock} - t_{DFF} - t_{setup}
    \label{eq:timing-constaints}
\end{equation}

Failure to meet these constraints results in the output DFF's value to be corrupted. Given $t_{DFF}$ and $t_{setup}$ are constant, two things can cause these errors: either the clock frequency being too high, or the combinational logic being too slow. The latter notably occurs when the system's supply voltage is insufficient, as the switching time of transistors is proportional to their supply voltage.

% --------------------------------------------------------------------

\section{Internal Energy-based Attacks} 
\label{sec:attacks}

As detailed in Section~\ref{sec:physical-attacks}, physical access to the victim device enables potent power and clock-based attack scenarios, which have been studied and utilized against secure microcontrollers. Recently, several studies have shown that similar attacks can be executed internally using software-accessible energy management mechanisms against complex \acp{CPU}. This shift is relatively new, with such attacks first demonstrated in 2015 for \acp{SCA}~\cite{yanStudyPowerSide2015} and in 2017 for \acp{FIA}~\cite{tangCLKSCREWExposingPerils2017}. Consequently, they pose an emerging threat that may become increasingly dangerous with further discoveries and when combined with other attack vectors.

%----- MODIFIED
Sensitive assets embedded in \ac{TEE} are common targets for internal energy-based attacks in recent literature~\cite{tangCLKSCREWExposingPerils2017,lippPLATYPUSSoftwarebasedPower2021}. Earlier attacks often targeted untrusted programs~\cite{yanStudyPowerSide2015}. However, with the widespread adoption of \acp{TEE}, research has shifted toward a privileged attacker model where trusted programs and data are the targeted assets, which are protected from direct access by the rich \ac{OS}. Energy management mechanisms -- voltage and frequency regulators and sensors -- are shared across security domains, spanning both \acp{TEE} and \acp{REE}. An attacker with full control over the \ac{REE} can exploit these energy management mechanisms as a side channel to access trusted assets. The methods by which these mechanisms can facilitate \acp{FIA}, \acp{SCA}, and covert communication will be described in the following paragraphs.
%----- DEIFIDOM

\subsection{Attacker Model} \label{sec:attacker-model}

This section describes the attacker's objectives and capabilities for conducting internal energy-based attacks from within the system. These attacks employ methods similar to the physical attacks described in Section~\ref{sec:physical-attacks}: they monitor programs using external power sensors and/or manipulate their execution using external sources of glitches. However, they differ significantly in a crucial aspect: internal attacks utilize software-accessible energy management mechanisms, eliminating the need for the attacker to physically access the device. While traditional physical attacks involve the attacker utilizing off-chip equipment to extract information or manipulate the execution flow, internal attacks are entirely software-driven, thus expanding the threat model to include remote attackers. This significantly increases the attack surface. Moreover, these internal attacks are more cost-effective, carry no risk of damaging the targeted platform, and pose a serious potential for widespread exploitation~\cite{shuvoComprehensiveSurveyNonInvasive2023}. Typically, the attacker first identifies a vulnerability on their own copy of the target device. Then, malware is developed to exploit this vulnerability, carrying out the attack scenario defined in the previous phase. This malware can then be deployed for remote exploitation.

\subsubsection{Attacker's Objectives.}
The attacker's goal is to compromise key security properties of the \ac{TEE} and trusted applications/enclaves. Namely, the targeted security properties are 1)~data confidentiality (\textit{e.g.}, stealing cryptographic keys), 2)~integrity (\textit{e.g.}, altering processed data), 3)~availability (\textit{e.g.}, conducting Denial-of-Service attacks), or 4)~authenticity (\textit{e.g.}, loading a self-signed application into the \ac{TEE}). This assumption implies that the CPU core currently running the \ac{TEE} is the ultimate target. Therefore, energy-based attacks targeting \acp{FPGA}~\cite{krautterFPGAhammerRemoteVoltage2018} are outside the scope of this work. However, attacks using on-chip \acp{FPGA} to target the \ac{CPU}, such as \ac{FPGA}-to-\ac{CPU} undervolting \acp{FIA}~\cite{mahmoudFPGAtoCPUUndervoltingAttacks2022}, are included.

\subsubsection{Attacker's capabilities.}
Our assumptions regarding the resources and capabilities of an attacker for internal energy-based attacks are based on the \ac{TEE} \ac{PP}~\cite{globalplatformTEEProtectionProfile2020} defined by the GlobalPlatform standards organization. This document serves as the primary reference for describing and evaluating the security of a \ac{TEE} implementation. Attacker models are defined in its appendix. Our model is primarily derived from the \emph{hardware vulnerabilities exploited from software}~\cite[Annex A, Section A.4.1]{globalplatformTEEProtectionProfile2020} attack path. For energy-based attacks, the \ac{SCA} and \ac{FIA} attack paths defined in the \ac{PP} are also considered. Thus, the main resources and capabilities of our assumed attacker are the following:
\begin{itemize}
    \item Full control over the \ac{REE}, including the
    \ac{OS}, its drivers, and access to debugging tools.
    \item Comprehensive documentation of the targeted \ac{TEE} implementation, including its hardware resources and cryptographic schemes.
    \item The ability to load a malicious trusted program in an attack on a cryptographic application allows the attacker to invoke operations with a targeted key, obtaining thousands of measurements of those operations.
    \item The capability to control the \ac{REE} to minimize noise due to signal processing and manage calls to the \ac{TEE} in a \ac{SCA}.
    \item The use of side-channel analyses, including non-energy-related \acp{SCA} such as cache attacks, to find a trigger signal in a \ac{FIA}.
\end{itemize}

Attackers are remote; they do not have physical access or proximity to the victim's device but can execute malicious software on it. However, they may utilize physical equipment on a copy of the device for profiling purposes before conducting the actual attack. For example, physical sensors can be employed to establish a correlation between power consumption and data on their own device during the profiling phase. Subsequently, during the actual attack, the developed model can be exploited on the victim's device through a remote attack using software-accessible power sensors embedded in the chip.

\subsection{Fault injection attacks}~\label{sec:fault-attacks}
Internal energy-based \acp{FIA} rely on inducing a modification of the supply voltage and frequency on the target CPU. As described in Section~\ref{sec:timing-constraints}, abrupt changes in these parameters can result in clock glitches, leading to faults in processed data or instructions. Typically, an unstable state where faults occur is reached when the supply voltage is low and the clock frequency is high. Above a certain threshold, faults occur in critical parts of the system, which produces a reset. The gap in which faults occur may be thin, depending on the device. The attacker can induce a glitch by (i) transiently increasing the frequency to an unstable value, known as 'overclocking' attacks~\cite{tangCLKSCREWExposingPerils2017}, or (ii) reducing the power domain's supply voltage while maintaining a high clock frequency, referred to as \emph{undervolting} attacks~\cite{qiuVoltJockeyBreakingSGX2019}. Table~\ref{tab:fault-attacks} provides an overview of internal energy-based \acp{FIA} published recently, highlighting their distinctive characteristics. In this section, we provide an overview of these attacks and their limitations, starting with attacks that maliciously exploit DVFS mechanisms and then addressing the specific case of FPGA-to-CPU undervolting attacks.

\begin{table*}[!tb]
    \centering
    \caption{\small Overview of \acp{FIA} exploiting software-accessible energy management mechanisms.}
    \renewcommand{\arraystretch}{3}
    \small
    \begin{tabular}{|>{\RaggedRight\arraybackslash}p{2.25cm}|>{\RaggedRight\arraybackslash}p{4.5cm}|>{\RaggedRight\arraybackslash}p{7.5cm}|}
         \hline 
         \thead{Attack} & 
         \thead{Target platform (Target \ac{TEE})} & 
         \thead{Compromised security properties (assets)} \\
         \hline 
         \makecell{\emph{CLKSCREW}~\cite{tangCLKSCREWExposingPerils2017}} & 
         \makecell{Qualcomm Krait (Arm TrustZone)} &
         \makecell{Confidentiality (extract AES keys from the \ac{TEE}), \\Integrity (fault an RSA key during verification),\\Authenticity (load a self-signed application into TrustZone)} \\
         \hline 
         \makecell{\emph{VoltJockey}~\cite{qiuVoltJockeyBreakingSGX2019,qiuVoltJockeyBreachingTrustZone2019}} & 
         \makecell{(i) Qualcomm Krait (Arm TrustZone)\\(ii) Skylake CPUs (Intel \ac{SGX})} &
         \makecell{Confidentiality (extract AES keys from the TrustZone \& \ac{SGX}),\\Authenticity (load a self-signed application into TrustZone)}\\
         \hline 
         \makecell{\emph{Plundervolt}~\cite{murdockPlundervoltSoftwarebasedFault2020}} & 
         \makecell{Skylake CPUs (Intel \ac{SGX})} & 
         \makecell{Confidentiality (Extract AES \& RSA keys from the \ac{TEE}),\\Integrity (Out-of-Bounds memory access)} \\
         \hline 
         \makecell{\emph{V0ltpwn}~\cite{kenjarV0LTpwnAttackingX862020}} & 
         \makecell{Skylake CPUs (Intel \ac{SGX})} & 
         \makecell{Integrity (fault SHA-256)}\\
         \hline
         \makecell{Noubir \textit{et al.}~\cite{noubirMaliciousExploitationEnergy2020}} &
         \makecell{Exynos 5422 \& Kirin 960} & 
         \makecell{Availability (denial-of-service attack)} \\
         \hline
         \makecell{FPGA-to-CPU \\undervolting \\attacks~\cite{mahmoudFPGAtoCPUUndervoltingAttacks2022,mahmoudDFAultedAnalyzingExploiting2022,grossFPGANeedlePreciseRemote2023}} &
         \makecell{Heterogenous SoCs \\(CPU and programmable layer)} &
         \makecell{Integrity (fault multiplications and AES encryption),\\Confidentiality (recover AES secret keys)}\\
         \hline
    \end{tabular}
    \label{tab:fault-attacks}
\end{table*}

\subsubsection{DVFS Attacks.}

DVFS \acp{FIA} involve installing a malicious kernel module or driver that accesses the hardware voltage and frequency regulators. By independently adjusting these parameters, the attacker causes the device to operate beyond its specified limits. This type of attack represents an emerging threat first demonstrated
in~\cite{tangCLKSCREWExposingPerils2017}. In this study, attackers overclock the device to induce a transient clock glitch, which is used to compromise the security of \ac{QSEE}, an Arm TrustZone-based \ac{TEE}. Additionally, as discussed in Section~\ref{sec:dvfs}, in most \acp{DVFS} implementations, each core (or cluster) has its own frequency domain. Attackers exploit this feature to target a specific core (or cluster) without affecting their attack program. Through this method, the authors successfully extract secret AES keys via \ac{DFA}. Moreover, they demonstrate the loading of a self-signed application into the \ac{TEE} by disrupting RSA certificate verification. Subsequent research has shown that transient undervolting can achieve similar results~\cite{qiuVoltJockeyBreakingSGX2019} and that Intel \ac{SGX} can also be compromised through energy management mechanisms by directly manipulating the corresponding \ac{MSR}~\cite{murdockPlundervoltSoftwarebasedFault2020,qiuVoltJockeyBreakingSGX2019,kenjarV0LTpwnAttackingX862020}. As shown in Table~\ref{tab:fault-attacks}, the target platforms encompass a range of devices, from smartphone processors (\textit{e.g.}, Qualcomm Krait) to high-end Intel \acp{CPU}. Both Arm TrustZone and Intel \ac{SGX} are shown to be vulnerable to \ac{FIA}. These attacks illustrate various use cases, compromising different security properties of the victim device and targeting diverse assets.

These studies present powerful attack scenarios but also reveal numerous limitations. In DVFS attacks, achieving precise timing for glitch injection is not always feasible, depending on how quickly the victim device can switch between different \acp{OPP}. In~\cite{tangCLKSCREWExposingPerils2017}, attackers were able to induce a glitch within a \num{65000}-cycle time window, targeting a specific part of the victim program. However, in Intel architectures examined in~\cite{murdockPlundervoltSoftwarebasedFault2020,kenjarV0LTpwnAttackingX862020}, over \num{500000} instructions are executed between consecutive \ac{OPP} change requests, making transient glitch injection impractical. Attackers exploit the fact that certain instructions, such as multiplications~\cite{murdockPlundervoltSoftwarebasedFault2020} and vector operations~\cite{kenjarV0LTpwnAttackingX862020}, are more likely to cause faults, allowing them to target specific parts of a program to some extent. When the victim device does not exhibit such behaviour, supply voltage manipulation can still be employed for denial-of-service attacks~\cite{noubirMaliciousExploitationEnergy2020}. Therefore, a short \ac{OPP} switch delay is favourable for \acp{FIA}. Additionally, voltage regulators require significantly more time to change the supply voltage compared to \acp{PLL} for changing the operating frequency. Thus, undervolting attacks, while being reportedly more difficult to prevent, offer less transience and precision compared to overclocking.

Secondly, not all SoCs give the OS direct access to voltage and frequency regulators. For devices lacking an open-source \ac{OS} or detailed technical documentation, the corresponding registers remain unknown. The attacks on Intel \ac{SGX}~\cite{murdockPlundervoltSoftwarebasedFault2020,kenjarV0LTpwnAttackingX862020,qiuVoltJockeyBreakingSGX2019} were made possible by prior research efforts that uncovered the \ac{MSR} used for voltage control~\cite{murdockPlundervoltSoftwarebasedFault2020}.

Third, each hardware device possesses unique characteristics that can make fault injection via voltage and frequency control challenging, if not impossible. The unstable region that triggers faults without requiring a device reboot or freeze may be very narrow on some \acp{CPU}. With discrete voltage control precision (\textit{e.g.}, \qty{5}{\milli\volt} steps), reaching this region may prove unattainable, as observed in AMD Zen processors~\cite{rabichSoftwarebasedUndervoltingFaults2020}.

\subsubsection{FPGA-to-CPU undervolting attacks.}
Recently, the use of programmable layers such as \acp{FPGA} to fault the CPU in heterogeneous platforms has been studied. This follows earlier studies which showed that FPGAs can be used to synthesise power-hungry circuits which provoke significant voltage drops when their activation is toggled at the resonance frequency of the \ac{PDN}~\cite{krautterFPGAhammerRemoteVoltage2018}. This vulnerability has mostly been studied in the context of multi-tenant cloud FPGAs for FPGA-to-FPGA attacks. However, the voltage drops affects all components in the same die, notably the CPU. A software attacker can leverage this vulnerability, along with the CPU's DVFS mechanisms, to perform an FPGA-to-CPU undervolting fault attack. These software-based FPGA-to-CPU attacks have been studied in~\cite{mahmoudFPGAtoCPUUndervoltingAttacks2022,grossFPGANeedlePreciseRemote2023,mahmoudDFAultedAnalyzingExploiting2022}. Although these works haven't yet explicitly targeted TEE assets, acting more as proof-of-concept attacks on bare-metal setups, they could be used in future work for remote attacks on \acp{TEE} since they target the underlying CPU. In~\cite{grossFPGANeedlePreciseRemote2023}, it is shown that this method can achieve precise fault injection, with fault models that were not observed in CPU-to-CPU undervolting attacks, namely instruction skips and data during transfer from DDR memory to caches. It is observed in~\cite{mahmoudDFAultedAnalyzingExploiting2022} that FPGA-induced attacks offer substantial improvement in timing precision compared to DVFS attacks. FPGA-to-CPU attacks have been used to successfully setup \ac{DFA} on an \ac{AES} encryption occuring on the CPU~\cite{grossFPGANeedlePreciseRemote2023,mahmoudDFAultedAnalyzingExploiting2022}. Thus, these attacks provide new fault models that could be exploited, together with the previously described DVFS attack, to jeopardize TEE security properties.

\subsection{Side-Channel and Covert-Channel Attacks} \label{sec:sca-attacks}
Side-channels can be exploited by attackers in two ways. First, sensitive programs can inadvertently reveal information through these channels. By collecting and processing this side-channel data in a \ac{SCA}, an attacker may retrieve sensitive assets. Second, an attacker may also covertly communicate information through these channels. Covert communication attacks involve a trojan and a spy; the trojan holds information, potentially sensitive data retrieved using another type of attack, that it secretly communicates to the spy through the side channel.

\subsubsection{Side-Channel Attacks.}

\begin{table*}[!htbp]
    \centering
    \captionsetup{justification=centering}
    \caption{\small Overview of \aclp{SCA} exploiting software-accessible energy management mechanisms. The resolution column defines the minimum time between 2 consecutive measurements.
    (P): With privileged access; (NP): With unprivileged access.
    }
    \renewcommand{\arraystretch}{1.5}
    \small
    \begin{tabular}{|>{\RaggedRight\arraybackslash}p{2.5cm}|>{\RaggedRight\arraybackslash}p{2cm}|>{\RaggedRight\arraybackslash}p{4.5cm}|>{\RaggedRight\arraybackslash}p{1.5cm}|>{\RaggedRight\arraybackslash}p{3.5cm}|}
         \hline 
         Attack & 
         Target Platform & 
         Attack Vector & 
         Time Resolution & 
         Compromised assets \\
         \hline
         Battery level reading attacks~\cite{michalevskyPowerSpyLocationTracking2015,yanStudyPowerSide2015,qinWebsiteFingerprintingPower2018,conf/iwsec/GiraudN23} & 
         Android smartphones & 
         Unprivileged battery level reading & 
         \qty{100}-\qty{175}{ms} & 
         Geo-localisation, running applications, website fingerprints. \\
         \hline 
         \emph{DF-SCA}~\cite{diptaDFSCADynamicFrequency2022} & 
         Linux-based systems (x86 and Arm) & 
         Unprivileged use of the \texttt{cpufreq} module & 
         \qty{10}{\milli\second} & Website fingerprints, keystrokes. \\
         \hline
         Attacks on RAPL counters~\cite{mantelHowSecureGreen2018,lippPLATYPUSSoftwarebasedPower2021,chowdhuryyPowSpectrePoweringSpeculation2024} & 
         x86 CPUs & 
         Energy consumption reading using RAPL counters & \qty{50}{\micro\second} (P) - \qty{1}{\milli\second} (NP)\footnotemark[2] & 
         AES keys from an \ac{SGX}-enclaved program, KASLR addresses. \\
         \hline
         Frequency throttling attacks~\cite{liuFrequencyThrottlingSideChannel2022,wangHertzbleedTurningPower2023} & 
         x86 CPUs, Arm SoCs, GPUs & 
         Manipulation and analysis of the frequency throttling mechanism & 
         --- &  
         Cryptographic keys (\textit{e.g.} AES) from enclaved programs, KASLR addresses, pixel sniffing. \\
         \hline
         \emph{Collide+Power}~\cite{koglerCollidePowerLeaking2023} & Any & Analysis of the power variation during replacement of attacker-controlled data by victim data, using direct measurement or frequency throttling. & --- & Any value in a co-located memory (e.g. caches).\\
         \hline
    \end{tabular}
    \label{tab:scas}
\end{table*}

Several studies demonstrate various malicious uses of energy management mechanisms targeting different assets. In the next paragraphs, we briefly describe the primary attack vectors that can be can be exploited for \acp{SCA}.

The first attack vector is a direct reading of energy-related metrics to carry. Attacks exploiting this vulnerability have evolved significantly over the last decade. The first published attacks exploiting this principle targeted unprivileged access to interfaces: battery status on Android phones~\cite{michalevskyPowerSpyLocationTracking2015,yanStudyPowerSide2015,qinWebsiteFingerprintingPower2018,conf/iwsec/GiraudN23}, and power consumption through the \texttt{cpufreq} module on Linux-based systems~\cite{diptaDFSCADynamicFrequency2022}. When accessed by an unprivileged user, these interfaces have a low refresh rate, ranging from ten to hundreds of milliseconds. Despite this low resolution, they can reveal information such as geolocation, website fingerprinting, or password guessing.

More recently, some studies describe how a privileged attacker can access energy consumption metrics, notably the \ac{RAPL} interface on x86 systems~\cite{mantelHowSecureGreen2018,lippPLATYPUSSoftwarebasedPower2021}. Privileged attackers benefit from a significantly higher refresh rate, with metrics being updated every millisecond or less. This enables more powerful attack scenarios, notably targeting  \ac{TEE} assets, such as retrieving AES keys from trusted programs.

The second attack vector is explored in~\cite{liuFrequencyThrottlingSideChannel2022,tanejaHotPixelsFrequency2023,wangHertzbleedTurningPower2023,wangDVFSFrequentlyLeaks2023}, where the attacker analyzes the behavior of frequency throttling automatically induced by \ac{DVFS} to maintain a balance between power, frequency, and temperature. These frequency adjustments are shown to be both instruction- and data-dependent~\cite{liuFrequencyThrottlingSideChannel2022}. An attacker can trigger them by subjecting the processor to heavy workloads. Moreover, a privileged user can manipulate the power budget threshold at which throttling occurs, facilitating the exploitation of this vulnerability. This methodology has been employed to compromise AES encryption within an \ac{SGX} enclave~\cite{liuFrequencyThrottlingSideChannel2022}. This vulnerability also extends to Arm \acp{SoC}, as demonstrated in~\cite{tanejaHotPixelsFrequency2023}, potentially making it exploitable against TrustZone as well. It is noteworthy that program execution time is directly proportional to clock frequency. Therefore, in attacks where clock frequency serves as the side channel, attackers can measure the execution time of a dummy program instead of directly accessing frequency metrics, as demonstrated in~\cite{wangHertzbleedTurningPower2023}. This renders frequency throttling attacks difficult to detect and prevent.

The third attack vector is used in~\cite{zhaoFPGABasedRemotePower2018}. It supposes that the attacker can access an FPGA that is using the same \ac{PDN} as the victim CPU. Power consumption variations across the PDN are mirrored as transient supply voltage drops for the FPGA, which induces delays in its combinatorial logic. Thus, the attacker can use the FPGA as a power monitor for the rest of the system. This method allows them to monitor the CPU's power consumption variations with higher temporal resolution than the previously mentioned methods. In~\cite{zhaoFPGABasedRemotePower2018}, it is shown that an attacker can break constant-time implementations of RSA keys using this technique. The potential use of this attack vector in other scenarios requiring greater precision is still to be determined in future work.

Finally, in~\cite{koglerCollidePowerLeaking2023}, Kogler \textit{et al.} demonstrate that these software-based power \acp{SCA} can reveal arbitrary values in shared memory components, such as caches. They observe that power consumption subtly varies when an attacker-controlled value is replaced by a victim program's value. This variation depends on the Hamming distance between the two values. By using two inverted attacker values and analyzing the impact of their eviction on power consumption, they amplify this subtle leakage signal. With this method, an attacker can leak single-bit differences in the victim data. This attack is \ac{CPU}-agnostic and can be executed with either direct measurement or frequency throttling to monitor power consumption.

Table~\ref{tab:scas} provides an overview of the software \acp{SCA} based on energy management mechanisms outlined in this section. It is evident that despite relying on low-resolution measurements, some attacks can still infer sensitive information. Moreover, internal energy-based \acp{SCA} appear to be more varied than \acp{FIA} in terms of potential attack scenarios. They are also evolving rapidly, with increasingly refined attack patterns emerging over the years.

\subsubsection{Covert Communication.}
Several studies have demonstrated that frequency and power management mechanisms can function as covert communication channels between a Trojan and a monitoring program. In these attacks, the attacker hijacks power management modules to establish a communication method based on frequency modulation. This modulation can be achieved in two primary ways. The first involves taking direct control of hardware frequency regulators~\cite{alagappanDFSCovertChannels2017,bossuetAdvancedCovertChannelsModern2023}. The second consists in subjecting the device to various workloads, resulting in voltage or frequency modulation through mechanisms such as \ac{AVS}, \ac{DVFS}, and current management techniques~\cite{haj-yahyaIChannelsExploitingCurrent2021,khatamifardPOWERTChannelsNovel2019}.

The throughput of these covert channels varies significantly, ranging from a few bits per second~\cite{alagappanDFSCovertChannels2017} to several megabytes per second~\cite{benhaniDVFSSecurityFailure2018}. This variability depends on several parameters. A privileged attacker with direct control over hardware regulators can change the frequency much more quickly than an unprivileged attacker who acts indirectly through workload adjustments. To detect these changes, utilizing programmable hardware, such as an \ac{FPGA} on the \ac{SoC}, proves to be significantly more efficient than software-based frequency measurement~\cite{bossuetAdvancedCovertChannelsModern2023}. Furthermore, the attack methodology itself influences throughput. For instance, disguising the communication within another protocol, such as \ac{ICMP} frames, restricts the maximum performance of the attack~\cite{yueConstructingTimingbasedCovert2014}. Finally, while existing research typically adopts a device-agnostic approach, customizing the attack method to the specific power management mechanisms of the targeted device can yield improved results~\cite{kalmbachTurboCCPracticalFrequencyBased2020}.

% --------------------------------------------------------------------

%------------- DISCUSSION
\section{Countermeasures} \label{sec:discussion}

As observed in Section~\ref{sec:attacks}, internal energy-based attacks, which encompass \acp{FIA}, \acp{SCA}, and covert communication, are potent, can be executed remotely, and are cost-effective as they do not require specialized equipment. Therefore, they pose a substantial threat to all commercial \ac{TEE} designs.

Despite the numerous demonstrations of these emerging attacks recently, many innovative TEE designs have yet to incorporate protections against them. This lack of protection is due to such vulnerabilities being perceived as hardware flaws~\cite{bahmaniCURESecurityArchitecture2021,costanSanctumMinimalHardware2016a} or because their mitigation is believed to be unrelated to \ac{TEE} protection mechanisms~\cite{leeKeystoneOpenFramework2020,schrammelSPEARVSecurePractical2023}. However, internal energy-based attacks do not exploit flaws in a single hardware block that can be remedied by strengthening its design. Instead, they capitalize on a fundamental vulnerability in the \ac{TEE} protection model: hardware regulators, accessible by the \ac{REE} without protection, inherently serve as a side channel. This side channel is particularly potent, yielding results comparable to those of energy and clock-based physical attacks, which are well-recognized and considered significant threats. Nonetheless, internal energy-based attacks cannot be classified in the same category as traditional physical attacks regarding the threat model. This is primarily because they do not utilize off-chip equipment but rather leverage components already embedded in the hardware device, thereby circumventing their primary objective of energy optimization. When no countermeasures are in place, any attacker with kernel-level privileges in the untrusted world (\textit{e.g.}, runs in the \ac{REE}), assumed in most \ac{TEE} adversary models, can execute this type of attack against an application running within the trusted world. Consequently, addressing this vulnerability at its root necessitates designing \ac{TEE} software and hardware mechanisms to either eliminate or render this side channel unusable for potential attackers.

Nevertheless, the most widely used \ac{TEE} implementations, Intel \ac{SGX} and Arm TrustZone, embed mitigations against the \ac{FIA} and some \acp{SCA} outlined in Section~\ref{sec:attacks}, although they hinder the use of power management mechanisms. Moreover, several studies have proposed proof-of-concept countermeasures against energy-based \acp{FIA}~\cite{koglerMinefieldSoftwareonlyProtection2023,mishraPlugYourVolt2023,zhangBlacklistCoreMachineLearning2018}. These proof-of-concept studies explore various approaches to mitigate energy-based attacks but have not yet been implemented in real-world \acp{TEE}. Finally, some well-established methods traditionally employed to counter hardware attacks could be adapted against internal energy-based attacks, given the similarity in their methodologies. For instance, masking against \acp{SCA} and redundancy against \acp{FIA}~\cite{taoSoftwareCountermeasuresDVFS2023}. All of these avenues offer potential inspiration for next-generation \ac{TEE} designs to incorporate tailored countermeasures to protect against energy-based attacks. 

In this section, we provide an overview and insights into existing approaches to counter internal energy-based attacks. First, we classify the countermeasures into three categories and provide the criteria we use to evaluate their potential in section~\ref{sec:countermeasures-classification}. We then examine existing implementations and proof-of-concept countermeasures against CPU-to-CPU energy-based fault attacks in Section~\ref{sec:fias-countermeasures}, and \acp{SCA} in Section~\ref{sec:scas-countermeasures}, highlighting the limits of each approach. We discuss the specific case of FPGA-to-CPU attacks in Section~\ref{sec:fpga-cpu-countermeasures}. Finally, we give a brief conclusions on countermeasures against internal energy-based attacks in Section~\ref{sec:countermeasure-conclusion}.

%---- NEW
\subsection{Countermeasures classification} \label{sec:countermeasures-classification}

\begin{figure*}[tb]
    \centering
    \includegraphics[width=0.7\linewidth]{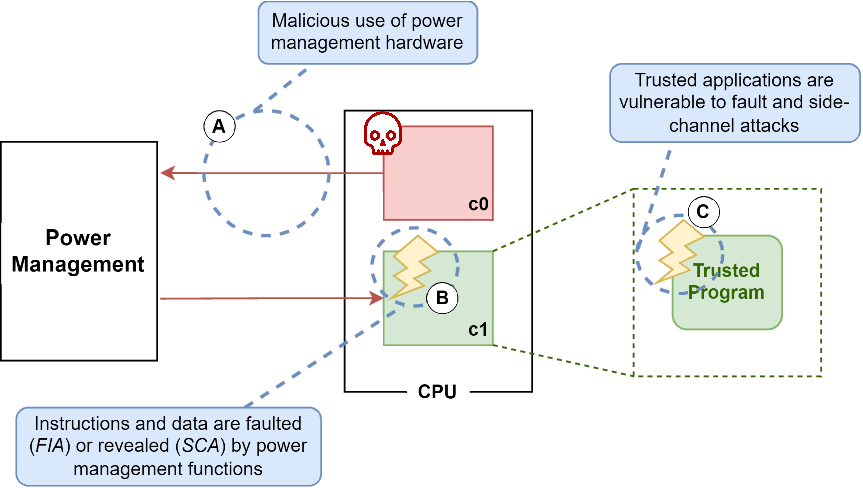}
    \caption{\small Main steps and components involved in internal energy-based attacks.}
    \label{fig:attack-steps}
\end{figure*}

Our classification of countermeasures against energy-based attacks is based on the main components involved in these attacks and their interrelations. The attack path is similar for both \acp{FIA} and \acp{SCA}. Attacks are facilitated by three vulnerabilities represented in Figure~\ref{fig:attack-steps}. To prevent energy-based attacks (both \acp{FIA} and \acp{SCA}), a countermeasure must mitigate at least one of them. Based on this observation, countermeasures can be classified into one of three classes, each corresponding to a vulnerability.

\begin{enumerate}[start=1,label={(\bfseries \Alph*):}]
    \item Attackers, using their root privilege in the \ac{REE}, can exploit power management hardware; whether to manipulate the supply voltage and frequency in an \ac{FIA} or to monitor energy-related metrics in an \ac{SCA}. This includes accessing software power management interfaces, or using devious means such as frequency throttling mechanisms and \acp{FPGA} to sense power consumption or to induce a voltage drop in the CPU. We define countermeasures as belonging to class \textbf{(A)} if their approach denies the attacker access to the power management hardware. Existing countermeasures in this category usually operate at software or firmware level.
    \item Power management mechanisms are directly linked to the target processor. First, they can provoke glitches if they don't supply the correct voltage and frequency. Second, the energy-related metrics that they collect (\emph{e.g.}, real-time power consumption, temperature) reveal information on which data are currently being processed, and how. Indeed, power consumption is data and instruction-dependent.  We define countermeasures as belonging to class \textbf{(B)} if they aim to prevent power management mechanisms from damaging the victim processor, \textit{i.e.}, by preventing faults due to incorrect voltage and frequency supply and stopping data and instructions from leaking through the power side channel. This can be achieved either by modifying power management mechanisms or by hardening the processor. Most of these countermeasures require hardware modifications.
    \item The victim's trusted program is sensitive to what happens on the underlying processor. If the processor faults, the program will encounter unexpected behaviour. Plus, its algorithm and secret values are processed by the leaky processor. We define countermeasures as belonging to class \textbf{(C)} if their objective is to harden trusted programs against the considered attack, for instance, by making them fault-tolerant or by designing a constant-power implementation of their algorithm. Therefore, this category of solutions operates at the level of the sensitive application.
\end{enumerate}

In the following sections, we use \textbf{(F)} to designate countermeasures approaches against energy-based fault attacks and \textbf{(S)} for those against side-channel attacks; followed by their class \textbf{(A)}, \textbf{(B)}, or \textbf{(C)}.

\paragraph{Criteria for assessing a countermeasure's potential.}
Besides the classification based on the previous paragraph, we use the following in our analysis to evaluate the potential of a countermeasure. The first criterion is the effectiveness of the countermeasure in negating the considered attacks. Indeed, while some solutions deal with the root cause of the vulnerability, effectively preventing the attack from happening, others focus on mitigating the consequences of an attack, or on reducing the attack's efficiency. Finally, it is important to note that some countermeasures have not been fully implemented, or have not been tested against actual attacks, making it difficult to guarantee and to evaluate their efficiency in practice.
A second criterion is the complexity of implementing and deploying the proposed solution. Depending on the chosen approach, a countermeasure can be more or less difficult to implement on the considered systems. Software solutions are practical in that regard because they can be deployed on existing devices. Meanwhile, countermeasures that require heavy hardware modifications are more suited for devices that have a security-oriented design, such as embedded systems used in critical applications.
Finally, the overhead required by the countermeasure needs to be considered. Countermeasures may affect the efficiency of the system, causing performance overhead for some programs, increasing their code size, and or the system power consumption.

In the following sections, we describe in depth the existing countermeasures against energy-based attacks, classify them and compare them based on the aforementioned criteria. First, we delve into countermeasures against \acp{FIA}, and second we study countermeasures against side-channel and covert-channel attacks.

\subsection{\acp{FIA} countermeasures} \label{sec:fias-countermeasures}
\begin {figure*}[tb]
\resizebox{\linewidth}{!}{
\begin{tikzpicture}[node distance=1.7cm]
    \node(base) [cat_rectangle] {DVFS \acp{FIA} Countermeasures};
    \coordinate[below of=base] (c_row1);
    
    \node(cat_1) [cat_rectangle, left of=c_row1, xshift=-6cm] {\textbf{(F-A)}: Prohibit control of the \ac{OPP} by the attacker};
    \node(cat_2) [cat_rectangle, below of=base,] {\textbf{(F-B)}: Prevent power management to fault the victim processor};
    \node(cat_3) [cat_rectangle, right of=c_row1, xshift=6cm] {\textbf{(F-C)}: Make trusted programs resistant to faults};
    
    \node(cat_1_1) [counter_rectangle, below of=cat_1,xshift=-0cm] {\textbf{(F-A1)}: Removal of software access to voltage regulators};

    \node(cat_1_ref1) [reference_rectangle, below of=cat_1_1, yshift=-0.5cm] {Intel \& Arm's official recommendations~\cite{armPowerPerformanceManagement2019,intelIntel64IA322023}};

    \node(cat_2_1) [counter_rectangle, below of=cat_2, xshift=-4.5cm] {\textbf{(F-B1)}: Reduce the accuracy of voltage \& frequency regulators};
    \node(cat_2_2) [counter_rectangle, below of=cat_2, xshift=-1.5cm] {\textbf{(F-B2)}: Avoid using unstable OPPs};
    \node(cat_2_3) [counter_rectangle, below of=cat_2, xshift=1.5cm] {\textbf{(F-B3)}: Maintain separate regulators across security boundaries};
    \node(cat_2_4) [counter_rectangle, below of=cat_2, xshift=4.5cm] {\textbf{(F-B4)}: Modify the CPU to increase its resilience to energy-based faults};

    \node(cat_2_ref1) [reference_rectangle, below of=cat_2_2, xshift=-3.7cm, yshift=-0.5cm] {Dedicated co-processor handling OPP change requests~\cite{zhangBlacklistCoreMachineLearning2018}};
    \node(cat_2_ref2) [reference_rectangle, below of=cat_2_2, xshift=-1.2cm, yshift=-0.5cm] {Monitor the current OPP from the Rich OS's kernel~\cite{mishraPlugYourVolt2023}};
    \node(cat_2_ref3) [reference_rectangle, below of=cat_2_2, xshift=1.2cm, yshift=-0.5cm] {Add a conservative OPP set along with the normal one~\cite{juffingerSUITSecureUndervolting2024}};
    \node(cat_2_ref4) [reference_rectangle, below of=cat_2_2, xshift=3.7cm, yshift=-0.5cm] {Reduce faultable instructions' critical path~\cite{juffingerSUITSecureUndervolting2024}};

    \node(cat_3_1) [counter_rectangle, below of=cat_3, ] {\textbf{(F-C1)}: Strengthen trusted programs against faults};
    \node(cat_3_ref1) [reference_rectangle, below of=cat_3_1, xshift=-3cm, yshift=-0.5cm] {Insert trap instructions to catch faults in trusted programs~\cite{koglerMinefieldSoftwareonlyProtection2023}};
    \node(cat_3_ref2) [reference_rectangle, below of=cat_3_1, yshift=-0.5cm] {Apply well-known software countermeasures against \acp{FIA}~\cite{taoSoftwareCountermeasuresDVFS2023}};

    \draw [arrow] (base) -| (cat_1);
    \draw [arrow] (base) -- (cat_2);
    \draw [arrow] (base) -| (cat_3);
    
    \draw [arrow] (cat_1) -- (cat_1_1);
    \draw [arrow] (cat_1_1) -- (cat_1_ref1);
    
    \draw [arrow] (cat_2) -| (cat_2_1);
    \draw [arrow] (cat_2) -| (cat_2_2);
    \draw [arrow] (cat_2) -| (cat_2_3);
    \draw [arrow] (cat_2) -| (cat_2_4);
    \draw [arrow] (cat_2_2) -- (cat_2_ref1.north east);
    \draw [arrow] (cat_2_2) -- (cat_2_ref2);
    \draw [arrow] (cat_2_2) -- (cat_2_ref3);
    \draw [arrow] (cat_2_4) -- (cat_2_ref4);
    
    \draw [arrow] (cat_3) -- (cat_3_1);
    \draw [arrow] (cat_3_1) -- (cat_3_ref1);
    \draw [arrow] (cat_3_1) -- (cat_3_ref2);

\end{tikzpicture}
}
%\end{adjustbox}

\caption{\small Existing and potential approaches for countermeasures against DVFS \acp{FIA}.}
\label{fig:countermeasures-faults}

\end{figure*}

The existing literature has proposed many approaches to counter or mitigate energy-based \acp{FIA}. Some of them have been implemented in commercial \acp{TEE}, or in proof-of-concepts described in publications and patents. These countermeasures greatly vary in their objectives, effectiveness, implementation abstraction level and overhead. They can consist of modifications in the device's software, hardware, or a combination of both. Given this great variety of countermeasures, we propose to summarize, analyse and compare them. 

Figure~\ref{fig:countermeasures-faults} gives a global overview of existing countermeasures approaches and their implementation based on the classification given in Section~\ref{sec:countermeasures-classification}. This graph shows that, while several countermeasures implementations rely on a similar principle, some potential approaches are still unexplored. In the next paragraph, we describe them more in-depth and compare their objectives and their potential.

\countermeasureSubsubsection{\textbf{(F-A)}: Prohibit control of power management by the attacker.}

\countermeasureParagraph{\textbf{(F-A1)}: Removing software access to regulators.}
This approach consists in disabling any software access to voltage regulators, even from the privileged rich OS. Indeed, to execute a \ac{FIA} as described in Section~\ref{sec:attacks}, the attacker loads a malicious kernel module that directly accesses hardware regulators. Therefore, this solution disarms the attacker and effectively prevents \ac{DVFS} \acp{FIA}. This countermeasure is recommended by manufacturers for \ac{TEE} implementations that have been targeted in this manner~\cite{armPowerPerformanceManagement2019,murdockPlundervoltSoftwarebasedFault2020}. More precisely, Arm recommends prohibiting the kernel from having \emph{direct and independent control of clock and voltage} and suggests performing regular checks on the requested \ac{OPP} using a trusted entity such as firmware~\cite{armPowerPerformanceManagement2019}. To our knowledge, no public document indicates that vendors (\emph{e.g.}, Qualcomm, Samsung, etc.) have implemented this countermeasure in their \ac{TEE}. Furthermore, in response to undervolting attacks on \ac{SGX}, Intel has disabled access to the voltage control \ac{MSR} when \ac{SGX} is enabled. This solely requires a firmware update, which is included in the \ac{TCB} attestation. An enclave can request this attestation to verify the effectiveness of the update. Thus, this solution is simple to deploy and highly practical for massively used \acp{TEE}.

However, this countermeasure runs counter to the primary objective of power optimization techniques. We assert that it may not be sustainable in the long term. Indeed, with this countermeasure, voltage management cannot be performed by the Rich OS and must be delegated to another component, such as hardware power management subsystems. No software control over voltage scaling implies that the system can only use the few \acp{OPP} defined by the manufacturer. This vendor-prescribed \acp {OPP} may diverge significantly from a device's actual operating limits~\cite{zhangBlacklistCoreMachineLearning2018}. It also precludes the ability to adjust them dynamically for precise control over the device's power consumption. In~\cite{juffingerSUITSecureUndervolting2024}, the effects of undervolting the device under these vendor-prescribed \acp{OPP} (while still staying within the operating limits) are studied. It is shown that this slight undervolting can bring significant improvements: 11.0\% efficiency gains, up to 20.8\% if the software's energy awareness is improved as well.

In summary, this solution is simple to implement and fully prevent DVFS-based \acp{FIA}. Plus, it does not add any performance overhead to the system. However, it impedes the utilization of \ac{DVFS} to its fullest extent for enhancing performance, minimizing energy consumption, and optimizing temperature. This is crucial in resource-constrained embedded devices, as well as in servers where effective power and temperature management are vital. Furthermore, it does not address the underlying vulnerability: there may be additional undocumented methods to manipulate voltage and frequency from the kernel~\cite{murdockPlundervoltSoftwarebasedFault2020}, including manipulating FPGAs to cause voltage drops on heterogeneous platforms~\cite{mahmoudFPGAtoCPUUndervoltingAttacks2022}. As a result, other countermeasures have been explored in the literature.

\countermeasureSubsubsection{(F-B): Prevent power management mechanisms to fault the target processor.}
\countermeasureParagraph{\textbf{(F-B1)}: Reduce the accuracy of regulators.}
As stated in Section~\ref{sec:fault-attacks}, the accuracy of voltage and frequency regulators directly affects the attacker's control over the induced glitch. First, if the voltage and frequency steps are too large, then the attacker may not be able to find an unstable operating area that would trigger a fault. Second, if delays required to switch the voltage and frequency from one step to another are too long, then the attacker loses precision over the fault, \textit{i.e.}, loses the ability to target a specific part of the victim program. The impact of tweaking these parameters on the feasibility of DVFS \acp{FIA} has not yet been studied. On the downside, this prevents energy-aware designers from using undervolting and similar techniques to their full potential for optimising efficiency. In addition, it goes against the general trend in the development of power management mechanisms towards faster and more precise control of the device's supply voltage and frequency.

\countermeasureParagraph{\textbf{(F-B2)}: Enforce operating limits to avoid using unstable \acp{OPP}.}

\acp{FIA} exploiting power management mechanisms rely on glitches occurring when the processor operates beyond its specified operating limits. Therefore, enforcing these limits in hardware would render such attacks impractical. However, implementing this in reality may pose challenges. Firstly, determining the actual operating limits of a device is only feasible after a post-manufacturing testing phase, making it practically difficult to enforce hardware-enforced hard limits~\cite{tangCLKSCREWExposingPerils2017}. Secondly, various factors contribute to the fact that even two similar devices manufactured on the same wafer may have different limits, such as variations and uncertainties during the manufacturing process, temperature fluctuations, and ageing effects. As a result, either a broad security margin must be applied, which restricts the optimal utilization of \ac{DVFS}, or the limits must be dynamically updated.

Therefore, this approach is more complex than \textbf{(F-A1)}, which disables software control over the \ac{OPP} altogether. However, it is less detrimental to energy efficiency, as long as the enforced range is less restrictive than the \acp{OPP} defined by the manufacturer. This range must strike a balance between safety (by being far enough away from the operating limits to prevent faults) and energy efficiency (by being as close to the limits as possible, allowing the OS to optimize energy consumption through undervolting). The performance overhead and ease of implementation of the countermeasures under consideration are also important factors to consider.

Several works propose countermeasure implementations that follow this approach. They first make a characterization of the device's stability depending on their supply voltage and frequency, by submitting it to various \acp{OPP}. This characterization is then used to determine a reasonable range, which is then enforced by software or hardware means.

In~\cite{mishraPlugYourVolt2023}, a kernel module polls the voltage and frequency registers to ensure that they do not cross the stability threshold. Although this module can be unloaded by an attacker who controls the rich \ac{OS}, an attestation can be used to verify its integrity. This approach is purely software and simple to implement. Plus, the authors observe that it introduces a low-performance overhead: on average 0.28\% performance loss over the \texttt{SPEC2017} benchmarks. However, the authors do not indicate how to dynamically update the threshold to account for the effects of ageing. Besides, it is unclear whether this countermeasure fully prevents DVFS \acp{FIA}. For instance, a fault could strike between two polls made by the kernel module.

In~\cite{zhangBlacklistCoreMachineLearning2018}, a hardware-level approach is proposed. This appears to make the countermeasure robust, as it is unlikely that an attacker could compromise the integrity of a separate hardware component. It relies on a lightweight machine-learning model, trained based on the previously mentioned device characterization. This model is used to predict whether an \ac{OPP} is safe to use or not. It runs on dedicated "blacklist-core" coprocessors, one per \ac{CPU} core, independent of both the \ac{CPU} and the \ac{PMIC}. The coprocessors intercept \ac{OPP} change requests from the CPU and only apply them if they are predicted to be stable. To account for the effects of temperature and ageing, the restrictiveness of the blacklist core is dynamically controlled by a decision boundary. It is updated when a fault occurs during runtime. The implementation of this countermeasure requires hardware changes, \textit{i.e.}, to add blacklist cores on the \ac{SoC} and to route calls to the \ac{PMIC} to them. It also requires software changes in the \ac{TEE} trusted code, \textit{e.g.}, to implement a behaviour for the decision boundary. Deploying this countermeasure is therefore more difficult than purely software solutions, as it requires the involvement of various actors, from the hardware integrator to the software developer. Regarding performance, the authors point out that the blacklist-core adds a negligible overhead to frequency and voltage change requests. One of these cores consumes \qty{10.5}{\milli\watt} at \qty{250}{\mega\hertz}. This static overhead in the device's energy performance should be weighed against the benefits of this countermeasure compared to simply restricting the usable operating points to a vendor-defined discrete state, using Intel \emph{SpeedShift} technology, for instance~\cite{schweikhardtDFSMixedCriticality2022}.

In~\cite{juffingerSUITSecureUndervolting2024}, Juffinger \textit{et al.}, present a hardware-software co-design introducing a second set of \acp{OPP} to the device. This supplementary set offers more aggressive energy savings compared to the original one, featuring lower supply voltages for the same operating frequencies while maintaining high performance. It is based on the observation that some instructions, \textit{e.g.}, multiplications and vector operations, are more likely to fault than others when x86 systems are submitted to undervolting. The OS is modified to switch to the conservative \ac{OPP} set when executing such instructions. The authors of this work point out that it is not primarily intended as a countermeasure against DVFS attacks, but as a tool to safely improve energy efficiency.
Nevertheless, we believe that this approach could potentially be used in an actual countermeasure with a few modifications. It can be implemented solely by software changes. The main implementation challenge is to choose the component responsible for the \ac{OPP} curve: it cannot be the attacker-controlled kernel, but it must control each instruction to account for its faultiness. 
Hardware-controlled \ac{OPP} curve changes, as in Intel's \textit{SpeedShift} technology, could also be explored. Another challenge is that such a countermeasure could degrade the performance of the system, which has to stall during each \ac{OPP} curve switch. A voltage change can take up to a millisecond on some devices. Juffinger \textit{et al.}, propose to mitigate this risk by introducing a time limit, which prevents curve switches from being too frequent. They also note that this method gives better energy savings when programs are optimized to avoid using faultable instructions.

\countermeasureParagraph{\textbf{(F-B3)}: Maintain separate hardware regulators.}

Another approach is to use separate regulators so that they are not shared between trusted and untrusted programs, depending on whether a \ac{CPU} core is in a secure or non-secure state~\cite{tangCLKSCREWExposingPerils2017}. Implementing such a solution poses several challenges. First, maintaining multiple physical regulators per core may incur significant costs. Second, at the software level, access to the regulators must be restricted based on the core and execution environment of the program issuing the request. This necessitates a comprehensive power management solution across different software layers in the trusted world, which could introduce performance overhead and enlarge the size of the \ac{TCB}. Reserving one or several \ac{CPU} cores exclusively for trusted programs, equipped with dedicated hardware regulators, would simplify the implementation of this countermeasure, as suggested in~\cite{qiuVoltJockeyBreachingTrustZone2019}. However, this approach would essentially resemble using a \ac{TPM} or a coprocessor. As demonstrated in Section~\ref{sec:tees}, this contradicts the goals of \acp{TEE}, which strive to offer isolation while utilizing the same hardware resources for both trusted and untrusted programs. 

Overall, this approach to prevent energy-based \acp{FIA} poses significant challenges to implement while respecting the \ac{TEE} design philosophy. Meanwhile, it doesn't offer any obvious advantages over the \textbf{(F-B2)} approach, for which some implementations seem capable of countering the vast majority, if not all, \ac{DVFS} error attacks. Furthermore, the design of a proof-of-concept demonstrating the feasibility and effectiveness of this approach remains an open challenge: to the best of our knowledge, no publicly available work has done so.

\countermeasureParagraph{\textbf{(F-B4)}: Increase the resilience of the processor to power-management-induced faults.}
As shown in Section~\ref{sec:timing-constraints}, clock glitches and voltage glitches occur because the processor instructions' critical paths get too large compared to the clock frequency. Therefore, relaxing some instructions' critical paths can prevent such glitches from occurring. This method is explored in~\cite{juffingerSUITSecureUndervolting2024}, in which, as mentioned in \textbf{(F-B2)}, a second set of \acp{OPP} is added along the default, conservative one. On x86 systems, undervolting provokes specific types of instructions to fault, notably multiplications, which are very frequently used in many types of programs. To avoid having to switch to the conservative \ac{OPP} curve at each multiplication, Juffinger \textit{et al.}, propose to make multiplication resilient to undervolting faults instead, by relaxing their critical path. As a result, each multiplication takes one additional clock cycle to complete. This does not significantly affect performance because of instruction pipelining. In summary, this approach involves heavy hardware modifications, directly into the CPU's arithmetic logic units. Therefore, it would be impractical to apply it to every instruction. However, as demonstrated in~\cite{juffingerSUITSecureUndervolting2024}, it can be an efficient tool when combined with other approaches.

Other possibilities, not limited to DVFS \acp{FIA}, can be derived from existing hardware countermeasures against faults in general. For instance, adding redundancy to operations can detect and mitigate all types of faults. At the hardware level, there are many possibilities to implement this. For instance, by duplicating the entire instruction stream. Alternatively, methods based on control-flow integrity enforcement can be explored~\cite{wernerProtectingRISCVProcessors2019}. Moreover, leveraging the RISC-V \ac{ISA} opens up opportunities for hardware and cross-layer countermeasures that operate at both, the compiler and hardware levels~\cite{laurentCrosslayerAnalysisSoftware2019,michellandLowlevelFaultModeling2024,yuceFAMEFaultattackAware2016}. As of now, there have been no published attempts to utilize these methods to mitigate internal energy-based attacks. However, they can be used to protect the device against them, as they do for other types of \acp{FIA}. This type of approach requires heavy modifications and may induce significant overhead. They are suitable for devices with high-security requirements that need to protect against a variety of threats, including but not limited to energy-based attacks.

\countermeasureSubsubsection{(F-C): Harden trusted programs and hardware against faults.}

Countermeasures outlined in the preceding paragraphs effectively prevent attackers from injecting faults into trusted programs, thereby thwarting all known \ac{DVFS} \acp{FIA}. However, in this paragraph, we delve into alternative approaches aimed at securing trusted programs against these attacks. These strategies involve modifying trusted programs themselves to render them resilient to faults. While these countermeasures do not preempt \ac{DVFS} attacks, they either detect faults upon occurrence and rectify their effects or render faults impractical for exploitation by attackers. Although these countermeasures typically introduce more performance overhead than the aforementioned approaches, this overhead is confined to TEE programs, with little impact on the rest of the system. Another notable distinction is that this type of countermeasure assumes that developers of trusted software bear the responsibility for safeguarding their programs from potential attackers. Consequently, they are only pertinent in cases where \ac{TEE} vendors have not furnished lower-level mitigations against \acp{FIA} from the outset.

A countermeasure proposed in~\cite{koglerMinefieldSoftwareonlyProtection2023} detects undervolting \acp{FIA} using a compiler extension. It involves inserting \emph{trap} instructions into the victim code. As mentioned in Section~\ref{sec:attacks}, undervolting Intel processors increases the likelihood of certain instructions being faulted (multiplications, vector operations). However, this countermeasure introduces significant overhead in terms of both execution time and code size of the protected program. For instance, in the tested scenarios, it resulted in a performance overhead of \qty{148.4}{\percent} and a code size overhead ranging from \qtyrange{50}{150}{\percent} to mitigate \qty{99}{\percent} of \ac{DVFS} \acp{FIA}. Additionally, this approach is tailored specifically for undervolting-based attacks on x86 Intel platforms and may not generalize well to other architectures.

Other methods, that are non-specific to energy-based attacks, can be used to secure trusted programs. Indeed, since \acp{FIA} have been studied for decades, many well-known software techniques can be used to secure programs against faults in general. Redundancy and error detection codes, such as parity checks, are commonly utilized to both detect and mitigate the effects of faults. This can be achieved through software means by duplicating instructions at compilation, manually implementing redundancy checks in the code software, or even duplicating the entire encryption or decryption process, although the latter incurs significant overheads~\cite{shuvoComprehensiveSurveyNonInvasive2023}. Another approach is a software implementation of infection, wherein a completely erroneous cypher is produced in the event of a fault during encryption, rendering the output cypher unusable for attackers. Tao et al., explore the potential use of these three countermeasures (parity codes, redundancy-based methods and ineffective computation) against \ac{DVFS} \acp{FIA} targeting AES encryption in~\cite{taoSoftwareCountermeasuresDVFS2023}. They determine that the most cost-effective countermeasure, in terms of both performance and code size overheads, is temporal redundancy (\textit{i.e.}, running the encryption process twice serially), with overheads of \qty{34.18}{\percent} and \qty{12.04}{\percent} in performance and code size, respectively. However, they do not empirically assess the robustness of this countermeasure against actual overclocking or undervolting attacks. In~\cite{huangInstructionVulnerabilityTest2019}, Huang \textit{et al.}, propose to substitute the most \textit{faultable} instructions with \textit{safer} ones. Some instructions cannot be replaced, and for some others, the substitutes do not provide satisfying protection against DVFS \acp{FIA} (the fault rate is reduced by about half or less).

In summary, countermeasures that modify trusted code to make it resilient against faults have the advantages of being easy to deploy and of leaving the untrusted part of the system unaltered. However, they usually add a lot of overhead to the protected code in terms of performance and code size. Besides, they do not prevent attacks from happening, but only mitigate their consequences. In that regard, other solutions \textbf{(F-A)} and \textbf{(F-B)} that deal with the root cause of the vulnerability will be more efficient most of the time. Thus, this type of solution seems more suitable when the system and the \ac{TEE} do not offer protection against energy-based \acp{FIA}. In that sense, it highlights the necessity of native, embedded countermeasures that deal with the vulnerability at its root.

\subsection{Side-Channel Attacks Countermeasures} \label{sec:scas-countermeasures}

\begin{figure*}[tb]
\centering
\begin{tikzpicture}[node distance=2.5cm]
    \node(base) [cat_rectangle] {Internal Power-Management-Based \acp{SCA} Countermeasures};
    \coordinate[below of=base] (c_row1);
    
    \node(cat_1) [cat_rectangle, left of=c_row1, xshift=-3cm, yshift=0.5cm] {\textbf{(S-A)}: Prohibit the attacker from exploiting power management mechanisms.};
    \node(cat_2) [cat_rectangle, below of=base, text width=4cm, yshift=0.5cm] {\textbf{(S-B)}: Prevent power management mechanisms to reveal the processor's data and instructions under processing.};
    \node(cat_3) [cat_rectangle, right of=c_row1, xshift=3cm, yshift=0.5cm] {\textbf{(S-C)}: Make trusted applications power-independent.};

    \node(cat_1_1) [counter_rectangle, below of=cat_1] {\textbf{(S-A1)}: Forbid untrusted programs to access vulnerable power management interfaces~\cite{liuFrequencyThrottlingSideChannel2022}.};

    \node(cat_2_1) [counter_rectangle, below of=cat_2, xshift=-1.5cm] {\textbf{(S-B1)}: Scramble power metrics gathered from software~\cite{intelRunningAveragePower2022}.};
    \node(cat_2_2) [counter_rectangle, below of=cat_2, xshift=1.5cm] {\textbf{(S-B2)}: Reduce the precision and refresh rate of power-related metrics gathered from software.};

    \node(cat_3_1) [counter_rectangle, below of=cat_3, ] {\textbf{(S-C1)}: Traditional techniques: masking, threshold implementations, key refreshing~\cite{demulderProtectingRISCVSideChannel2019,liuFrequencyThrottlingSideChannel2022}.};

    \draw [arrow] (base) -| (cat_1);
    \draw [arrow] (base) -- (cat_2);
    \draw [arrow] (base) -| (cat_3);
    
    \draw [arrow] (cat_1) -- (cat_1_1);
    
    \draw [arrow] (cat_2) -- (cat_2_1);
    \draw [arrow] (cat_2) -- (cat_2_2);
    
    \draw [arrow] (cat_3) -- (cat_3_1);
\end{tikzpicture}

\caption{\small Mind-Map: Existing and potential approaches and implementations for countermeasures against power management based side-channel and covert-channel attacks.}
\label{fig:countermeasures-scas}

\end{figure*}

Unlike \acp{FIA}, \acp{SCA} are passive and thus more challenging to detect and thwart. Because they exploit the same vulnerabilities as covert channel attacks, it is reasonable to assume that similar countermeasures could mitigate both types of attacks. In the following, the existing approaches to counter internal energy-based \acp{SCA} are studied according to the classification given in Section~\ref{sec:countermeasures-classification}. This is represented, as well as existing implementations of these approaches, in Figure~\ref{fig:countermeasures-scas}. It can be observed that compared to energy-based \acp{FIA}, few countermeasures have been proposed and implemented for \acp{SCA}. Yet, many more \acp{SCA} attack scenarios have been demonstrated. This gap in the literature suggests that countermeasures for internal energy-based \acp{SCA} is a promising avenue for future research.

\countermeasureSubsubsection{(S-A): Prohibit malicious access to power management interfaces.}
To counter the first category of side-channel attacks, based on direct reading of the power consumption or battery level, these interfaces may simply be made inaccessible to all untrusted programs. However, said interfaces are extensively used by energy-aware programs and frameworks. Thus, simply deactivating them would be hurtful to the energy efficiency of the system. We advocate for the use of less restrictive countermeasures such as those presented in \textbf{(S-B)} and \textbf{(S-C)}.

For the other category of energy-management-related \acp{SCA}, which employs frequency throttling as the side channel~\cite{wangHertzbleedTurningPower2023,liuFrequencyThrottlingSideChannel2022,tanejaHotPixelsFrequency2023,wangDVFSFrequentlyLeaks2023}, attackers can facilitate the attack by altering the power or temperature threshold at which throttling occurs~\cite{liuFrequencyThrottlingSideChannel2022}. This can be effectively mitigated using \acp{TEE} by restricting access to corresponding configuration registers or files to trusted programs only. Such restriction compels attackers to increase the system workload to induce throttling, which is less practical. However, this comes in contradiction with one of the main reasons for which this throttling mechanism exists in the first place: power clamping, \textit{i.e.}, allowing the user to define an energy budget over a given period of time~\cite{rountreeDVFSFirstLook2012}.

We also point out that power management mechanisms may be maliciously used in much more ways for \acp{SCA} than for \acp{FIA}. Creative and unexpected attack vectors can be exploited. For instance, using an embedded programmable hardware component as a precise power sensor~\cite{bossuetAdvancedCovertChannelsModern2023}, or using thermal measurements instead of power traces~\cite{mishraTooHotHandle2024}. Therefore, finding all possible vulnerabilities and forbidding their access may be challenging for \ac{SoC} designers. There may always be new ways to exploit power management mechanisms.

\countermeasureSubsubsection{(S-B): Prevent power management from leaking information on the processed data and instructions.}
To thwart attacks like \emph{Platypus}~\cite{lippPLATYPUSSoftwarebasedPower2021}, which derive information from direct readings of instantaneous power consumption or frequency, several solutions can make the side channel challenging for attackers to exploit. Intel's approach involves scrambling software-accessible power traces when \ac{SGX} is enabled~\cite{intelRunningAveragePower2022}. Power traces sent directly to hardware components remain unaffected. The scrambling is achieved by randomizing the values sent to the software-accessible voltage \acp{MSR}. This allegedly makes this internal power side-channel unusable for the attacker, but it also induces a variation in the energy reporting of about 5-10\%, which can potentially harm some energy-aware programs which need precise data on the energy consumption. This imprecision is to be put in perspective with the inherent approximation of the estimation-based RAPL energy reporting~\cite{hackenbergPowerMeasurementTechniques2013}. However, recent works show that Intel's countermeasure is not sufficient to counter power-based \acp{SCA}. In ~\cite{chowdhuryyPowSpectrePoweringSpeculation2024}, Chowdhurry \textit{et al.}, bypass Intel scrambling countermeasure by combining a power-based \acp{SCA} with a \emph{transient execution attack}. This category of attacks uses architectural vulnerabilities to replay the victim program's instructions. Thus, the attackers can gather thousands of power measurements for the targeted instructions. With enough of them, Intel's noise-based countermeasure becomes inefficient.

The same method could potentially be transposed for frequency throttling attacks. Randomly tweaking the operating frequency may prevent the attacker from inferring information from measurements. A similar approach has been proposed against physical power attacks in~\cite{shengqiyangPowerAttackResistant2005}. With a finite set of frequency \acp{OPP}, it still may be possible for the attacker to infer information even when noise is added to the operating frequency.

Another potential approach is to reduce the granularity of software-accessible counters and embedded sensors to hinder energy-based \acp{SCA}. However, as shown in part~\ref{sec:sca-attacks}, attackers may still infer sensitive information even with a low-resolution channel (\textit{e.g.}, one measurement every \qty{10}{\milli\second}).

In summary, this type of countermeasures are interesting and leads to mitigating energy-based \acp{SCA}. Indeed, they are less restrictive for power management than simply deactivating these interfaces, while being allegedly efficient to prevent this type of attack~\cite{hackenbergPowerMeasurementTechniques2013}. They do not totally prevent the attacker from gathering information but offer a statistical-based countermeasure. Therefore, there may still be ways to exploit such scrambled or imprecise data. In addition, the application of these methods for frequency throttling attacks is still to be studied. Finally, this approach has the same drawback as for those of the \textbf{(S-A)} category: given the variety of ways to exploit power management mechanisms for \acp{SCA}, it may be challenging to implement a mitigation for all of them.

\countermeasureSubsubsection{(S-C): Prevent trusted programs from revealing information.}
Traditional algorithm-level countermeasures against power analysis (both physical and remote) can be employed to increase the difficulty of exploiting these attacks. This type of countermeasure concentrates on making the attacks harder to exploit by obscuring results or concealing sensitive information. These solutions encompass masking, threshold implementations~\cite{nikovaThresholdImplementationsSideChannel2006}, and key refreshing, which may be implemented in trusted programs to shield them against both types of attacks~\cite{demulderProtectingRISCVSideChannel2019,liuFrequencyThrottlingSideChannel2022}. Their potential in the specific case of internal power-management-based \acp{SCA} has yet to be assessed. Given that this type of countermeasure requires to modify the algorithms themselves, it can imply some overhead, contrary to previously mentioned approaches. However, it has the advantage of leaving power management mechanisms themselves unaltered, which benefits energy-aware programs. This type of countermeasure also tends to focus solely on cryptographic algorithms, whereas \acp{TEE} encompass a much wider variety of programs.

\subsection{Countermeasures against FPGA-to-CPU attacks} \label{sec:fpga-cpu-countermeasures}
As described in Section~\ref{sec:attacks}, several works showcase how an attacker can use an \ac{FPGA} to exploit power management hardware and attack CPU programs, both in \acp{SCA} and fault attacks. These attacks extend from previous research, which focused on FPGA-to-FPGA attacks in a multi-tenant scenarios. To the best of our knowledge, no countermeasure exists specifically for energy-based FPGA-to-CPU attacks. However, several methods have been proposed to counter FPGA-to-FPGA attacks~\cite{krautterMitigatingElectricallevelAttacks2019,laFPGADefenderMaliciousSelfoscillator2020,nassarLoopBreakerDisablingInterconnects2021,provelengiosMitigatingVoltageAttacks2021}. These countermeasures aims to prevent an attacker from making malicious usage of FPGAs for precise power sensing and for inducing voltage drops. In that regard, they also prevent FPGA-to-CPU attacks.

Two main approaches have been proposed in the literature. The first consists in analysing bitstreams prior to deploying them on the FPGA to potentially detect malicious circuits~\cite{krautterMitigatingElectricallevelAttacks2019,laFPGADefenderMaliciousSelfoscillator2020}, as an antivirus would do for software programs. It is designed to forbid both the deployment of power sensors to carry out an SCA, and power-hungry circuits that can be used to induce a voltage drop. The analyser looks for patterns that are known to be used in such attacks. This includes searching primitives such as Ring Oscillators or unusual connections between the clock and data paths among others, and carrying out timing analysis to detect runtime behaviour that would provoke high current variations~\cite{krautterMitigatingElectricallevelAttacks2019}. This type of countermeasure has two main shortcomings: first, it prohibits the use of legitimate IPs that use potentially harmful circuits for benign purposes. Second, because it is based on analysis of existing attacks, it is not effective against as yet unknown attack schemes~\cite{nassarLoopBreakerDisablingInterconnects2021}. Therefore, it can't completely prevent undervolting attacks.

The second approach is to detect voltage drops at runtime and prevent them by disabling the malicious part of the FPGA. The main challenge with these countermeasures is to disable the malicious circuit fast enough to prevent the attack. Contrary to the previous approach, this is helpful only against undervolting attacks and not FPGA-based \acp{SCA}. Voltage drop detection can be achieved using voltage fluctuation sensors, similar to the ones used in \acp{SCA}. To neutralise the attacker's circuits, methods based on clock gating~\cite{provelengiosMitigatingVoltageAttacks2021}, switching all interconnects to a high impedance state~\cite{nassarLoopBreakerDisablingInterconnects2021} and partial reconfiguration~\cite{kajolAHDLAMNewMitigation2023} have been studied. At present, such countermeasures can prevent attacks leading to a crash, but not fault attacks, since faults occur faster than crashes during a voltage drop~\cite{nassarLoopBreakerDisablingInterconnects2021}. However, FPGA-induced fault attacks are successfully detected with this method, which may allow the system to trigger other mitigations.

\subsection{Summary on existing countermeasures against internal energy-based attacks} \label{sec:countermeasure-conclusion}

As demonstrated in the preceding paragraphs, the focus on internal energy-based attacks has predominantly centred on securing processors against \acp{FIA}. However, the solutions adopted by manufacturers — namely, restricting direct kernel control over \ac{OPP} and relying on governor-based \ac{DVFS} management — limit the optimal utilization of \ac{DVFS}, resulting in energy inefficiency. Furthermore, the emergence of throttling-based attacks~\cite{wangHertzbleedTurningPower2023} underscores that governor-based \ac{DVFS} also presents an exploitable side-channel for attackers. This approach fails to address the underlying cause of these attacks: the shared and accessible nature of energy management mechanisms by both the \ac{TEE} and the \ac{REE}, inherently constituting a side-channel. Nevertheless, the implementation of separate regulators for the two domains has not been achieved yet and may prove costly. Recent works proposing specific countermeasures against energy-based fault injection attacks~\cite{koglerMinefieldSoftwareonlyProtection2023,zhangBlacklistCoreMachineLearning2018,mishraPlugYourVolt2023} offer promising avenues for securing \acp{TEE}, albeit highlighting the substantial challenges in terms of overhead and implementation complexity. Additionally, we posit that leveraging features of modern \ac{TEE} designs, such as the exclusive assignment of hardware peripherals~\cite{bahmaniCURESecurityArchitecture2021} or attestation, could enhance countermeasure effectiveness. Besides, this allows the \ac{TEE} to stay the sole trusted actor in the system, thereby centralizing the \ac{TCB}.

Meanwhile, while specific countermeasures against DVFS \acp{FIA} proposed to date~\cite{koglerMinefieldSoftwareonlyProtection2023,zhangBlacklistCoreMachineLearning2018} show promise, they await implementation in actual hardware \ac{TEE} technologies. Lastly, few published work has proposed countermeasures against energy-based software \acp{SCA} beyond hardening the algorithms under attack or restricting access to their interfaces, akin to \acp{FIA}. The design and implementation of innovative and efficient countermeasures against power-management-based internal \acp{SCA} remains an open challenge.

% --------------------------------------------------------------------

\section{Conclusion}
\label{sec:conclu}

In this article, we conducted a thorough literature review of a novel category of software-induced hardware attacks, focusing on the malicious exploitation of power management mechanisms. We also examined existing countermeasures aimed at mitigating these attacks and explored potential strategies for integrating them into \acp{TEE}. Our analysis underscores the considerable challenges inherent in safeguarding \ac{TEE} implementations against energy-based attacks, highlighting this as a promising avenue for future research. Our key conclusions can be summarized as follows.

Internal energy-based attacks, including \acp{FIA}, \acp{SCA}, and covert communications, pose a significant threat to \ac{TEE} implementations. This evolving threat landscape demands attention as attackers continue to develop more sophisticated and potent attack vectors. Despite the increasing prominence of these attacks, previous academic work on hardware \ac{TEE} designs has largely overlooked them. Therefore, it is imperative that next-generation \ac{TEE} technologies integrate robust countermeasures to address this evolving threat landscape.

Emerging countermeasures against internal energy-based \acp{FIA}, as evidenced in recent literature~\cite{koglerMinefieldSoftwareonlyProtection2023,zhangBlacklistCoreMachineLearning2018}, offer promising avenues to secure \acp{TEE} against these threats without overly constraining their power management mechanisms. Nonetheless, the practical implementation and testing of these countermeasures, either in commercial \acp{TEE} or in academic RISC-V based \ac{TEE} designs, require further evaluation. Additionally, the exploration of mitigation strategies for energy-management-based \acp{SCA} remains an open area for future research.

% --------------------------------------------------------------------

\section*{Acknowledgments}

This research is supported by the French Research Agency (ANR) under the CoPhyTEE JCJC project contract ANR-23-CE39-0003-01.

% --------------------------------------------------------------------

\bibliographystyle{ACM-Reference-Format}
\bibliography{References_v2}

\vfill

\end{document}